\documentclass[useAMS,usenatbib]{mn2e}
\usepackage{epsf,amsfonts,amsmath,amssymb}
\usepackage{graphicx}
\usepackage{ulem}
\usepackage{mathrsfs}
\usepackage{color}

\def\G{\Gamma}

\title[Energies of GRB blast waves]{Energies of GRB blast waves and prompt efficiencies as implied by modeling of X-ray and GeV afterglows}

\author[P. Beniamini et al.]{Paz Beniamini$^{1}$
\thanks{E-mail:paz.beniamini@gmail.com}, Lara Nava$^{1}$, Rodolfo Barniol Duran$^{2}$ and Tsvi Piran$^{1}$\\
$^{1}$Racah Institute for Physics, The Hebrew University, Jerusalem, 91904, Israel\\
$^{2}$Department of Physics and Astronomy, Purdue University, 525 Northwestern Avenue, West Lafayette, IN 47907, USA}

\begin{document}

\date{Accepted ... Received ...; in original form ...}

\pagerange{\pageref{firstpage}--\pageref{lastpage}} \pubyear{2002}

\maketitle

\label{firstpage}

\begin{abstract}
We consider a sample of ten GRBs with long lasting ($\gtrsim10^2\rm\,sec$) emission detected by Fermi/LAT and for which X-ray data around $1\,$day are also available.
We assume that both the X-rays and the GeV emission are produced by electrons accelerated at the external forward shock, and show that the X-ray and the GeV fluxes lead to very different estimates of the initial kinetic energy of the blast wave.
The energy estimated from GeV is on average $\sim50$ times larger than the one estimated from X-rays. We model the data (accounting also for optical detections around $1\,$day, if available) to unveil the reason for this discrepancy
and find that good modelling within the forward shock model is always possible and leads to two possibilities: either the X-ray emitting electrons (unlike the GeV emitting electrons) are in the slow cooling regime or ii) the X-ray synchrotron flux is strongly suppressed by Compton cooling, whereas, due to the Klein-Nishina suppression,
this effect is much smaller at GeV energies. In both cases the X-ray flux is no longer a robust proxy for the blast wave kinetic energy.
On average, both cases require weak magnetic fields ($10^{-6}\lesssim \epsilon_B \lesssim 10^{-3}$) and relatively large isotropic kinetic blast wave energies $10^{53}\rm\,erg<E_{0,kin}<10^{55}\rm\,erg$ corresponding to large lower limits on the collimated energies, in the range $10^{52}\rm\,erg<E_{\theta,kin}<5\times10^{52}\rm\,erg$ for an ISM environment with $n\sim 1\mbox{cm}^{-3}$ and $10^{52}\rm\,erg<E_{\theta,kin}<10^{53}\rm\,erg$ for a wind environment with $A_* \sim 1$.
These energies are larger than those estimated from the X-ray flux alone, and imply smaller inferred values of the prompt efficiency mechanism,
reducing the efficiency requirements on the still uncertain mechanism responsible for prompt emission.
\end{abstract}

\section{Introduction}
\label{Int}
Gamma-Ray Bursts (GRBs) have two phases. A short prompt phase in which the emission is predominantly in the sub-MeV range is followed by a multi-wavelength afterglow that can be observed up to several years after the prompt emission. 
According to the generally accepted internal-external model \citep{Sari(1997)} the 
prompt emission is thought to originate from internal energy dissipation that takes place within the relativistic outflow. Conversely, the afterglow radiation is thought to originate from interactions between the outflow and the external medium \citep{Rees(1992),Paczynski(1993),Meszaros(1997),Sari(1997)}.
While the origin of the afterglow radiation has been understood in its general properties of synchrotron emission \citep{Sari(1998)},
the nature of the prompt radiation has not yet been completely unveiled. Indeed, the nature of both the dissipation and the radiative mechanisms is still uncertain.
The composition of the outflow, its initial energy content, and the processes at the origin of the prompt radiation are among the most relevant open issues in GRB studies. 

A critical issue in this model is the ratio of energy involved in the first (prompt) and second (afterglow) phases. 
This ratio reflects the efficiency of the prompt radiation process, a very important clue on the nature of this mechanism.
The kinetic energy left in the blast wave after the prompt phase can be estimated from afterglow observations.
Assuming that the cooling frequency lies below the X-ray band, the X-ray luminosity provides a robust estimate of the energy stored in the accelerated electrons, which in turn is directly related to the kinetic energy of the outflow \citep{Kumar(2000),Freedman(2001)}.
Under this assumption, several studies have exploited both pre-Swift X-ray observations \citep{Berger(2003), LRZ(2004)}, and Swift X-ray observations \citep{Berger(2007),Nysewander(2009)}.
Most of these studies have inferred a relatively low kinetic energy, which implies quite a large prompt efficiency: 
$\epsilon_{\gamma}>0.5$.
The discovery of the X-ray plateaus in many of the Swift GRBs increased the severity of the efficiency problem. The X-ray flux at the beginning of the plateau phase (around 500 sec) is lower by a factor $\sim3$ as compared with the flux estimated by extrapolating backwards in time the observations at $\sim1\,$day
and therefore leads to an estimate of the kinetic energy lower by the same factor and to efficiencies of up to $90\%$ \citep{Granot(2006),Ioka(2006),Nousek(2006),Zhang(2007)}.

Internal shocks are the dominant internal dissipation process for matter dominated flows 
\citep{Narayan(1992),Rees(1994)}. Since their efficiency is rather low \citep{Kobayashi(1997),Daigne(1998),Beloborodov(2000),Kobayashi(2001),Guetta(2001)} it was expected that after the prompt phase most of the energy would remain as bulk kinetic energy of the blast wave. 
Alternative scenarios, such as magnetic reconnection \citep{Usov(1992)} (that require a magnetically dominated rather than a matter dominated outflow) may reach higher efficiencies, leaving less energy in the blast wave.
Thus the high efficiency, implied by the X-ray afterglow observations, is generally considered as a major problem for the internal shocks model and suggested that other mechanisms, such as magnetic reconnection take place. 

However, two implicit assumptions have been made when deriving these estimates: first it was assumed that the electrons emitting at X-rays are fast cooling (i.e. the X-ray is above all the typical synchrotron break frequencies) and second
the X-ray flux of these electrons is not suppressed by Synchrotron-Self Compton (SSC) losses \citep{Sari(2001), Fan(2006)}.
If either one of these assumptions is not satisfied the energy estimates based on the X-ray fluxes might be wrong.

Observations at higher energies could be helpful in constraining the location of the synchrotron cooling frequency and assess the importance of the SSC mechanism. 
GRBs can now be studied at energies between $0.1\,$GeV and  $300\,$GeV thanks to the Large Area Telescope (LAT) on board Fermi.
While only a small fraction of Fermi detected GRBs have also been detected by LAT,
some of these cases (about 10 events) are particularly interesting, since they show characteristics suggestive of an external shock origin for the GeV radiation:
first, the onset of these emission is delayed relative to the onset of the prompt sub-MeV emission \citep{Abdo a(2009)}; second the LAT component 
extends long after the prompt sub-MeV emission stops and third the flux of this long lasting component decays as a power-law in time.
Indeed, these observations are compatible with expectations from forward shock radiation 
\citep{KBD(2009),KBD(2010),Ghisellini(2010),Wang(2013),Nava(2014)}.

We can exploit the afterglow observations at high energies to address the questions of the determination of the synchrotron cooling frequency
and the importance of the SSC mechanism (which leads to a determination of the energy content of the fireball and the efficiency of the prompt mechanism).
We examine a sample of GRBs detected {\it both} by LAT and XRT, and use both the GeV and the X-ray fluxes to estimate the blast wave kinetic energy.
We show that the two estimates are inconsistent with each other. The energy inferred from the GeV emission is much larger than that estimated from the X-rays.
This can be explained if either i) the X-ray emitting electrons, unlike those emitting at GeV energies, are in the slow cooling regime,
or ii) if electrons radiating at X-rays are significantly cooled by SSC, while those radiating at GeV energies are in the Klein-Nishina (KN) regime and cool only by synchrotron.
In both scenarios the X-ray flux is no longer a proxy for the blast wave kinetic energy. 
We examine afterglow models for the GeV, X-ray and (when available) optical data in order to determine if one or both of the mentioned scenario represents a viable solution to the XRT/LAT apparent inconsistency.

The nature of the solution depends on the cooling frequency and on the value of the Compton parameter. Both depend strongly on the fraction of energy stored in the downstream magnetic field $\epsilon_B$ and on the density of the external medium $n$. Modelling of the data allows us, therefore,
to constrain these parameters. Recent works on GRBs with GeV emission \citep{KBD(2009),KBD(2010),Lemoine(2013)} and without it (\citealt{RBD(2014),Santana(2014)}; \citealt{Zhang(2015),Wang(2015)}) suggested a distribution for the $\epsilon_B$ parameter that extends down to very small values: $10^{-6}-10^{-7}$.
In this work we derive upper limits on $\epsilon_B$ as a function of the external medium density and find similar results: upper limits on $\epsilon_B$ derived form our sample range from $10^{-2}$ down to $10^{-6}$.

The paper is organized as follows. In \S \ref{Sample} we describe the sample of GRBs used in this paper. In \S \ref{efficiency}
we show that the kinetic energies of the blast waves inferred from GeV data are much higher than those inferred from the X-ray observations, and in turn the corresponding prompt efficiencies are much smaller.
We examine, in \S \ref{sec:model}, both scenarios and we provide analytic estimates of the parameters needed to explain the observations.
We present the numerical method used to model the data and describe the results of the detailed numerical modeling in \S \ref{sect:results}. Finally, in \S\ref{Conclusions} we summarize our conclusions.

\section{The Sample}
\label{Sample}
We consider a sample of GRBs detected both by Fermi-LAT and by Swift-XRT.
We included in our sample only those bursts for which the LAT emission lasted much longer than the prompt phase,
since we are interested in cases in which the LAT is most likely afterglow radiation from the external forward shock.
When available we use optical observations to further constrain the solution.
The final sample includes ten GRBs, listed in Table \ref{tbl:data}. 
The redshift has been firmly measured for nine of those.

To ensure that the measured fluxes are most likely originated at the forward shock with no reverse shock contribution
and to make a proper comparison with previous studies, we chose XRT observations that are as close
as possible to $\sim1\,$day. 
We select two observation times for each burst. The first such that it satisfies the following conditions: i)
they are as close as possible to one day, ii) subsequent to the end of any plateau phase in the light-curve that might be present (this phase is present only for one burst in the sample, GRB 090510) and iii) before the jet break time (when present).
The second observation time is chosen such that it is as far removed from the first observation as possible (in order to enable a good estimate of the light-curve power law index) while satisfying conditions ii) and iii).
Such late observations are not available for the LAT and we therefore use earlier epochs for the LAT measurements. 
For LAT we use two data points as far removed in time as possible given that they are after $T_{90}$ (the duration of the prompt emission as determined by the lower energy Fermi / GBM detector)
and provided that the flux measurement is still well constrained.

For the LAT fluxes we used values reported in the First Fermi/LAT GRB catalog \cite{ACK13} for nine of the bursts, while for GRB 130427A we used the results reported in \cite{ACK14}.
The X-ray fluxes are taken from the Swift/XRT GRB light curve repository\footnote{http://www.swift.ac.uk/xrt\_curves/} \citep{Evans(2007),Evans(2009)} and the optical fluxes are
collected from the literature. The complete list of LAT, X-ray and Optical fluxes and observations times is reported in Table \ref{tbl:data}.

\begin{table*}
\tiny
\begin{tabular}{l l l l l l l l l l l l l l l l}
\hline
Burst & $t_{GeV,1}$ &$F_{GeV,1}$ & $t_{GeV,2}$ &$F_{GeV,2}$& $t_{X,1}$ &$F_{X,1}$ & $t_{X,2}$ & $F_{X,2}$   &
$t_{opt,1}$ & $F_{opt,1}$  & $t_{opt,2}$  & $F_{opt,2}$ & $z$ & $t_{jet}$ & ref. \\
 & $10^{-3}$days & nJy & $10^{-3}$days & nJy & days & nJy & days & nJy & days & $\mu$Jy & days & $\mu$Jy & &days& \\
\hline
080916C & 4.6 & 4.9 & 1& 72 &  6.94 & 9.7 & 11.57 & 3.6 &1.39 & 5.5 &3.47 &1.5 & 4.35&$>15.3$ & \footnotemark[1]\\
090323  & 4.3 & 14& 1.6& 54&  2.495 & 23& 5.78 & 6.3 &1.85 & 14& 5.1 & 2.7 & 3.57 &$>10$& \footnotemark[2]\\
090328  & 14 & 1.2 & 2& 8.1 & 3.47 & 20 & 6.9 & 11 &1.63 & 25&2.6 & 11 & 0.73 &$>10$ & \footnotemark[2]\\
090510  & 1.3 & 16 & 0.1& 210 &  0.14 & 22 &0.062 & 100 &1.16 & 1.8 & 0.14 & 9 & 0.9 & $>0.75$ &\footnotemark[3]\\
090902B & 7.6 & 15 & 0.2& 400 & 0.928 & 21 & 2.38 &3.7 & 1.43 & 10 & 2.52 & 5.7& 1.822 & $20$&\footnotemark[4] \\
090926A  & 3 & 15 & 0.4& 42 &  2 & 80 &11.57 & 5.4 & 3 & 37& 6.1 & 14000& 2.1 &$10$ & \footnotemark[2]\\
100414  & 3.2 & 7.1 & 0.5& 66 &  2.3 & 7.6 & 7.17 & 0.33 & - & - & -& -& 1.37 &  $>7.4$&-\\
110625A & 5 & 38 & 3& 42 &  0.46 & 130 & 0.139& 1300& - & - & -& - & * & $>0.47$&-\\
110731A  & 3.5 & 0.75 & 0.2& 350 &  1.22 & 61& 3.84& 12 & 0.026& 70& 0.012 & 240& 2.83 & $>7.5$&\footnotemark[5]\\
130427A  & 63 & 2 & 3& 110 &  1.19 & 2600 & 11.54& 100 &1.15 & 130 & 0.22 & 1000  & 0.34 & $>180$&\footnotemark[6]\\
\hline
\end{tabular}
\caption{The GeV, X-ray and optical  fluxes and observation times  for bursts in the sample. 
The GeV data is taken from the first Fermi/LAT GRB catalog \citep{ACK13}, apart from 130427A, which is taken from \citep{ACK14}.
The X-ray data is taken from the Swift/XRT GRB light curve repository.
The jet break estimates for 090902B and 090926A are taken from \citealt{Cenko(2011). The lower limits on the jet breaks for the other bursts are taken from the latest available X-ray flux.}
The last column lists the references for the optical data:
$^1$\citealt{Greiner(2009)};
$^2$\citealt{Cenko(2011)};
$^3$\citealt{Nicuesa Guelbenzu(2012)};
$^4$\citealt{Pandey(2010)};
$^5$\citealt{Lemoine(2013)};
$^6$\citealt{Vestrand(2014)}.
*For GRB 110625A there is no measured
redshift and we use a typical value of $z=1$.}
\label{tbl:data}
\end{table*}

\section{Kinetic energy of the blast wave}
\label{efficiency}
We begin by re-visiting the estimates of the kinetic energy of the afterglow that are based on X-ray observations.
As in previous works, we assume for now that the X-ray radiation around one day is above the cooling frequency $\nu_c$ (the frequency above which electrons cool efficiently) and that Compton losses are negligible (i.e. $Y \ll 1$, where $Y$ is the Compton parameter). 
In this case the kinetic energy scales as: $E^{\rm x}_{0,k} \propto F_X^{4 \over 2+p} \epsilon_B^{2-p \over 2+p} \epsilon_e ^{4(1-p) \over 2+p} t_X^{3p-2 \over 2+p}$,
where $t_X$ is the time of X-ray observations and $F_X$ is the X-ray flux at that time (see Eq. \ref{eq:aboveC} below).
To estimate the kinetic energies, we use the data in Table \ref{tbl:data} (in particular we use $t_{X,1}$ and $F_{X,1}$ as well as $t_{GeV,1}$ and $F_{GeV,1}$) and assume that $\epsilon_e=0.1$, $p=2.5$ and $\epsilon_B=10^{-2}$. The results
depend only weakly on $p$ and $\epsilon_B$. The resulting energy estimates are similar to those found in previous studies
 \citep{Frail(2001),Panaitescu(2001a),Panaitescu(2001b),Granot(2006),Ioka(2006),Nousek(2006),Zhang(2007),Nysewander(2009),Cenko(2011)}.
In particular, they imply large gamma-ray prompt emission efficiencies: $\langle\epsilon_{\gamma,X}\rangle=0.87$ (see Table \ref{tbl:eff}), where $\epsilon_{\gamma} = E_{\gamma}/(E_{\gamma}+E_{0,kin})$, and the sub-index indicates whether X-ray (``X") or GeV fluxes (``GeV") are used to estimate it. The efficiency is as large as $0.99$ for GRB090902B.

We repeat the same calculation using the GeV fluxes, assuming that this high energy radiation is also produced by forward shock synchrotron. 
We use photons detected by LAT in the 0.1-10 GeV range and it is therefore expected that they are in the fast cooling regime of the spectrum. 
If this is the case and if $\epsilon_e$ and the prompt efficiency are narrowly distributed, then a strong correlation between GeV luminosity and $E_{\gamma}$ is expected \citep{Ghisellini(2010),Beniamini(2011)}. This has indeed been found by \cite{Nava(2014)} for a sample of 10
GRBs which almost completely overlap with the sample of bursts studied in the present work, strengthening the hypothesis that the high energy radiation is synchrotron radiation from the external forward shock at frequencies larger than $\nu_c$.
Compton cooling of the emitting electrons is suppressed by the KN lower cross section \citep{Wang(2010)}, that is relevant at such high energies.
Therefore, the GeV flux is expected to be a better proxy of the kinetic energy than the X-ray flux.

\begin{figure*}

\centering
\includegraphics[width=0.4\textwidth]{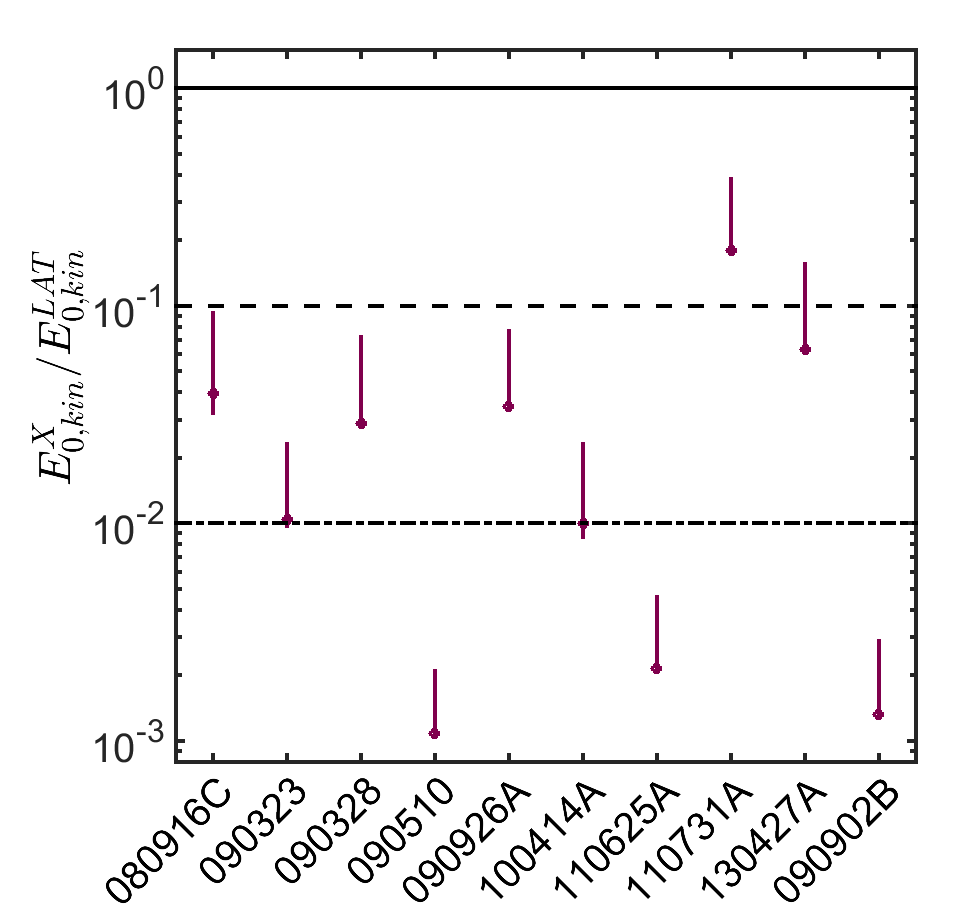}
\includegraphics[width=0.4\textwidth]{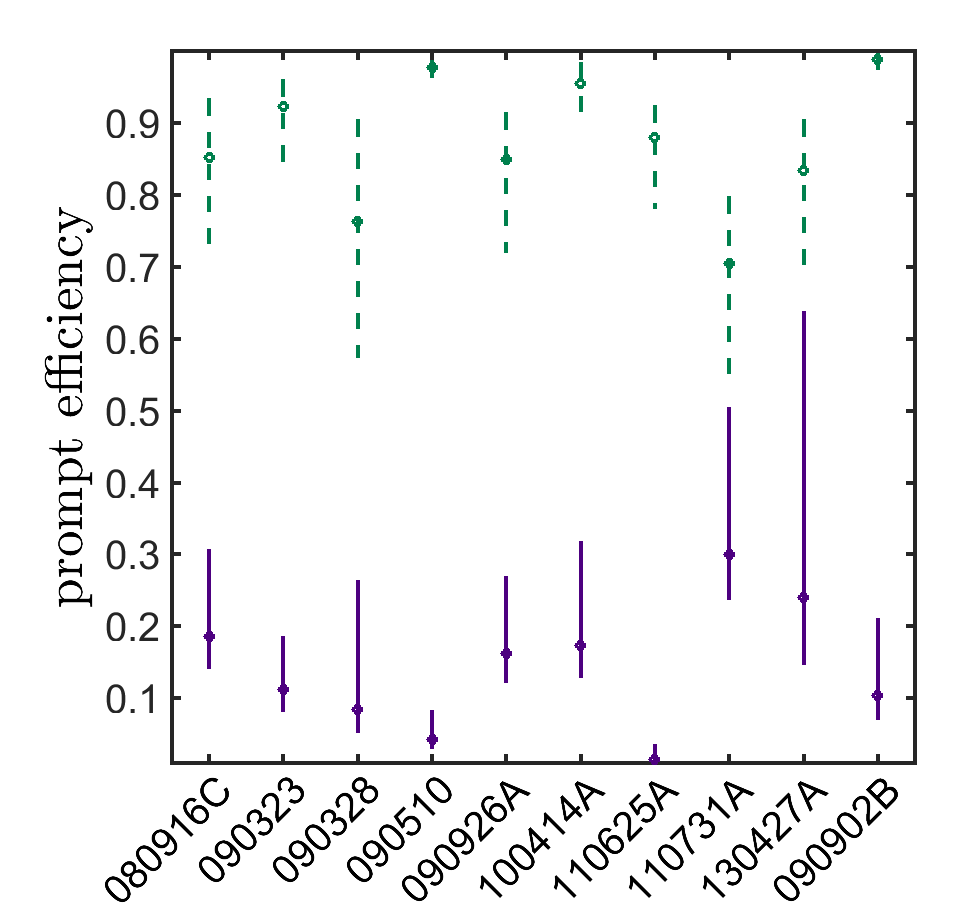}
\caption
{\small Left panel: ratio between the kinetic energy of the blast wave as estimated from X-rays ($E^{\scriptscriptstyle X}_{0,kin}$) and the same quantity as estimated from GeV flux ($E^{\scriptscriptstyle GeV}_{0,kin}$).
Both X-rays and GeV are assumed to be above $\nu_c$ at the time of measurement.
Note that the ratio of the estimates is independent of $\epsilon_B, \epsilon_e$ and $n$.
The solid line marks equal energy, the dashed line marks $E^{\scriptscriptstyle X}_{0,kin}=0.1E^{\scriptscriptstyle GeV}_{0,kin}$ and the dot-dashed line marks $E^{\scriptscriptstyle X}_{0,kin}=0.01E^{\scriptscriptstyle GeV}_{0,kin}$.
Errors on the ratio take into account possible radiative losses (that might reduce the energy content of the blast wave between the time of GeV
observations and the time of X-ray observations) and different values of the electron distribution power law index $p$ (from 2.1 to 2.8). Right panel:
efficiencies of the prompt phase of the GRBs in the sample assuming the kinetic energies from X-rays (dashed lines) and GeV (solid lines) shown in the left panel.}
\label{ExElat}
\end{figure*}

The kinetic energies $E^{\scriptscriptstyle GeV}_{0,kin}$ obtained from the GeV flux are much larger than kinetic energies $E^{\scriptscriptstyle X}_{0,kin}$ obtained from X-rays (see Table. \ref{tbl:eff}).
The inconsistency between the estimated kinetic energies can be better appreciated in terms of their ratios, shown in Fig. \ref{ExElat}. Typically $E^{\scriptscriptstyle GeV}_{0,kin}$ is at least ten times and up to a hundred times larger than $E^{\scriptscriptstyle X}_{0,kin}$.
The efficiencies inferred using the GeV observations are considerably lower than those obtained from the X-ray data: $\langle\epsilon_{\gamma,GeV}\rangle=0.14$, as compared to $\langle\epsilon_{\gamma,X}\rangle=0.87$ inferred from X-ray data.

This ratio, ${E^{\scriptscriptstyle X}_{0,kin}}/{E^{\scriptscriptstyle GeV}_{0,kin}}$, is independent of the assumptions made on $\epsilon_B, \epsilon_e$ (assuming these quantities do not change significantly between the time of the two observations), and it depends very weakly on $p$ through:
$(t_X/t_{GeV})^{3p-2 \over 2+p} (\nu_X/\nu_{GeV})^{2p \over 2+p}$ (where $\nu_X=1\,$keV is the frequency at which the X-ray flux
is obtained and $\nu_{GeV}=100\,$MeV is the equivalent frequency for GeV). 
We estimate the ratio ${E^{\scriptscriptstyle X}_{0,kin}}/{E^{\scriptscriptstyle GeV}_{0,kin}}$ for $p=2.5$ and incorporate the uncertainty on $p$ ($2.1 \lesssim p \lesssim 2.8$) in the error bars in the figure. 
The error bars also account for possible radiative losses that decrease the kinetic energy content of the fireball between the time of the GeV and the X-ray observations. 
To account for these losses, 
we follow \cite{nava13} to estimate the radiative losses for each burst assuming $\epsilon_e=0.1$ and a "fast cooling" regime (this is a conservative assumption: it assumes that all the energy dissipated
at the shock and transferred to the electrons, is radiated). 
It is clear from Fig. \ref{ExElat} that the variation of $p$ and possible radiative losses are not sufficient to explain the difference. 
The inconsistency implies that at least one of the naive assumptions was wrong. We turn in the next section to modeling of the afterglow that will enable us to examine this issue.

\section{Analytic estimates}
\label{sec:model}
We turn now to model the GeV and X-ray data in order to understand which one of the initial assumptions is not valid. 
To recall, these two assumptions were that either both the X-rays and the GeV emission are in the fast cooling regime and that neither emitting electrons are cooled by Inverse Compton. 
We envisage two possibilities.
In the first, ``SSC suppressed " scenario,
both the X-rays and the GeV photons are above $\nu_c$, but the X-ray flux is suppressed due to IC losses, while the GeV emitting electrons are in the KN regime and their flux is not suppressed \citep{Wang(2010)}.
If this is the case, and if Compton cooling is not accounted for properly,
then a small kinetic energy is inferred from the relatively low X-ray flux, while a higher kinetic energy is inferred from the high-frequency, unsuppressed, GeV flux. 
A second possibility is the ``slow cooling" scenario in which the X-ray band is below $\nu_c$.
In this case the X-ray flux depends strongly also on $\epsilon_B$ and on the external density and much higher energy 
is needed to produce a given flux than if the X-rays were above $\nu_c$. 
In both scenarios (SSC suppressed or slow cooling) the X-ray flux is no longer a good proxy for the kinetic energy of the blast wave.

\begin{table*}\small 
\begin{tabular}{ l l l l l l}
\hline
Burst & $E_{\gamma,54}$& $E^{\scriptscriptstyle X}_{0,kin,54}$ & $E^{\scriptscriptstyle GeV}_{0,kin,54}$ & $\epsilon_{\gamma,X}$ & $\epsilon_{\gamma,GeV}$ \\

\hline

080916C &3.48 & 0.6 (1.38)& 15.2 (20.1)& 0.85 (0.71)&0.19 (0.14)\\
090323  & 3.44 & 0.28 (0.61)& 27.4 (37.3)& 0.92 (0.85)&0.11 (0.08)\\
090328  & 0.1 & 0.03 (0.07)& 1.08 (1.71)& 0.77 (0.58)&0.08 (0.05)\\
090510  & 0.04 & 0.001 (0.002)&0.88 (1.12)& 0.98 (0.96)& 0.04 (0.03)\\
090902B & 2.53 & 0.03 (0.06) & 21.4 (32.1) & 0.99 (0.98)& 0.1 (0.07)\\
090926A  & 1.75 & 0.3 (0.67)& 9 (12.3)& 0.85 (0.72)&0.16 (0.12)\\
100414  & 0.49 & 0.023 (0.05)& 2.3 (3.2)& 0.95 (0.9)&0.17 (0.13)\\
110625A & 0.18 & 0.02 (0.05)& 11.4 (16.6)& 0.88 (0.78)&0.015 (0.01)\\
110731A  & 0.49 & 0.2 (0.4)& 1.1 (1.5)& 0.7 (0.54)&0.3 (0.24)\\
130427A  & 0.8 & 0.15 (0.35)& 2.5 (4.6)& 0.83 (0.7)&0.24  (0.15)\\
\hline
\end{tabular}
 \caption{Comparison of prompt gamma-ray energy \citep{ACK13} and derived kinetic energies using either X-ray or GeV data for the bursts in the sample. Column 2 lists the prompt isotropic equivalent gamma-ray energies of the bursts.
Columns 3,4 list the derived kinetic energies using the X-ray flux - at $t_{X,1}$ (see Table \ref{tbl:data}) and GeV flux - at $t_{GeV,1}$
(see Table \ref{tbl:data}) accordingly. Columns 5,6 list the derived efficiencies for the same data.
These values are estimated for $p=2.5$ and $\epsilon_B=10^{-2}$ in an ISM medium. They change only mildly for different
values of $p$ and $\epsilon_B$ and for a wind medium. The values in parenthesis correspond to the same quantities in case the outflow was affected by radiative losses.}
\label{tbl:eff}
\end{table*}

We begin by considering simplified analytic estimates to explore if and how the two proposed scenarios can explain the apparent kinetic energy inconsistency.
The analytic estimates allow us to understand what are the conditions that the free parameters (and in particular $\epsilon_B$ and $n$) must satisfy in order to
reconcile the kinetic energy estimates from both X-ray and the GeV observations.

In order to present analytic estimates we make the simplified assumption that the GeV radiation is always above $\nu_c$ and that due to KN suppression, Compton cooling does not affect the spectrum at these frequencies \citep{Wang(2010)}, i.e. $\nu_{GeV}>\nu_{KN}$, where $\nu_{KN}$ is the KN frequency, above which seed photons undergo KN suppressed IC scatterings, and is given by
$\nu_{KN}\approx\G m_e c^2/(h (1+z)\gamma_e)\leqslant \G m_e c^2/(h (1+z)min(\gamma_m,\gamma_c))$ (where $h$ is Planck's constant, $\Gamma$ is the bulk Lorentz factor,
$z$ is the cosmological redshift and $\gamma_e$ is the typical Lorentz factor of electrons in the co-moving frame which is always equal to or larger than the minimum Lorentz factor - 
either $\gamma_m$ for slow cooling or $\gamma_c$ for fast cooling). If $\gamma_m<\gamma_c$, the upper limit to the KN frequency is constant in time and depends only on $\epsilon_e$. For $\epsilon_e=0.1$, its value is around $\sim10^{18}$Hz, well below the LAT band. At early times $\gamma_m$ might be larger than $\gamma_c$, and $\nu_{KN}$ is larger that what estimated in the previous case, but still below the LAT range.
Therefore, in first approximation, the GeV flux is a proxy of the kinetic energy.
$E_{0,kin}$ can be derived from Eq. \ref{eq:aboveC} (see below), using $Y\!=\!Y_{GeV}\!\ll\!1$ $(Y_{GeV}$ is the Compton parameter for GeV emitting electrons). This assumption greatly simplifies the analytical expressions. The validity of this assumption is revisited in \S \ref{sect:results}, where we
numerically estimate $Y_{GeV}$, instead of assuming $Y_{GeV}\ll1$. We find that in any case its value is of order unity or smaller, and the approximation $Y_{GeV}\ll1$ that we use in this section is good enough to derive order
of magnitude estimates.
We fix the fraction of shocked energy in electrons to $\epsilon_e=0.1$.
We do this for four reasons. First, contrary to $\epsilon_B$, the values of $\epsilon_e$ from afterglow modeling do not vary over orders of magnitude and are often consistent with 0.1 \citep{Santana(2014)}. Second, this value is consistent with numerical simulations of shock acceleration \citep{Sironi(2011)}. Third, \cite{Nava(2014)} have shown (for a sample of 10 bursts that almost completely overlaps with the one used here)
that $\epsilon_e$ must be narrowly distributed and that GeV light curves are consistent with a value 0.1.
Fourth, since the radiated energy is limited to the energy in the electrons ($\epsilon_e E_{0,kin}$),
the inferred energy from the observed flux is larger for smaller values of $\epsilon_e$. Therefore it cannot be much lower than 0.1 in order to avoid un-physically 
large energy requirements. 
These considerations significantly limit the allowed range of this parameter and allow us to fix it without much loss of generality.
We are left with three free parameters: $\epsilon_B$, $p$, and $n$ (or $A_*\equiv A/(5\times10^{11}\mbox{g/cm})$, where $A$ is the
wind parameter defined by $n=(A/m_p)r^{-2}$ for a wind environment). 
We may therefore use the expression for the X-ray flux to obtain $\epsilon_B$ as a function of $n$ (or $A_*$), and $p$.

\begin{subsection}{SSC suppressed case}
\label{S:SSC}
If Compton losses are important, the flux above the cooling frequency scales
as $F_{\nu}(\nu>\nu_c)\propto E_{0,kin}^{(2+p)/4}(1+Y)^{-1}$. 
Therefore $\frac{E^{\scriptscriptstyle X}_{0,kin}}{E^{\scriptscriptstyle GeV}_{0,kin}} \propto (\frac{1+Y_X}{1+Y_{GeV}})^{4/(2+p)}$ ($Y_X$ is the Compton
parameter for the X-ray emitting electrons). The difference in energies can be explained as a difference in the respective Compton $Y$ parameters.
In the SSC suppressed case, the expression for the X-ray flux, $F_X$, is that of synchrotron above the cooling frequency \citep{Granot(2002)}:
\begin{equation} 
\label{eq:aboveC}
F_\nu\!=\!
\left\{ \!
  \begin{array}{l}
 0.855 (p-0.98) e^{1.95p}\ (1\!+\!z)^{2+p\over 4}\ \bar\epsilon_{e}^{\, \,p-1}\ \epsilon_B^{p-2\over 4} \\ \ E_{kin,52}^{2+p\over 4}
 \ t_{\rm days}^{2-3p\over 4}\ d_{L28}^{-2}\ (\frac{\nu}{\nu_{14}})^{-p/2}\ (1\!+\!Y)^{-1} \mbox{mJy,}\quad \mbox{for ISM}\\
 \\
 0.0381\ (7.11-p)\ e^{2.76p}\ (1\!+\!z)^{2+p \over 4}\ \bar\epsilon_{e}^{\, \, p-1}\ \epsilon_B^{p-2\over 4}\\ \ E_{kin,52}^{2+p\over 4}
\  t_{\rm days}^{2-3p\over 4}\ d_{L28}^{-2}\ (\frac{\nu}{\nu_{14}})^{-p/2}\ (1\!+\!Y)^{-1} \mbox{mJy,}\quad \mbox{for wind}\\
  \end{array} \right.
\end{equation}
where $\bar\epsilon_e \equiv \epsilon_e \frac{p-2}{p-1}$, $d_L$ is the luminosity distance, $t_{\rm days}$ is the time since explosion
in days, both $t$ and $\nu_{14}$ are in the observer frame
and we use the notation: $q_x = q/10^x$ in c.g.s. units here and elsewhere in the text
(the numerical coefficients of the ISM and wind cases in Eq. \ref{eq:aboveC} are not very different).
We assume that X-ray emitting electrons are in the Thomson regime, $\nu_m<\nu_c$ and $2<p<3$. In this regime, the Compton parameter $Y_X$ is given by (see Appendix \ref{ComptonY} for a derivation):
\begin{equation}
\label{Yx}
 Y_X=\frac{\epsilon_e}{\epsilon_B (3-p) (1+Y_X)}\bigg(\frac{\nu_m}{\nu_c}\bigg)^{p-2\over 2}.
\end{equation}
Note that this is an implicit equation, as $\nu_c$ also depends on $Y_X$ (see Eq. \ref{eq:nuc} below).
The typical frequencies, $\nu_m, \nu_c$ (where $\nu_m$ is the synchrotron frequency of the typical electron energy) are given by: 
\begin{equation} 
\label{eq:num}
\nu_m=
\left\{
  \begin{array}{l}
3.73\ (p-0.67)\  10^{15}\  (1+z)^{1/2}\ E_{kin,52}^{1/2}\ \\ \bar\epsilon_{e}^2\ \epsilon_B^{1/2}\ t_{\rm days}^{-3/2}\mbox{Hz,}\quad \mbox{for ISM}\\
\\
4.02\ (p-0.69)\ 10^{15}\ (1+z)^{1/2}\ E_{kin,52}^{1/2}\ \\ \bar\epsilon_{e}^2\ \epsilon_B^{1/2}\ t_{\rm days}^{-3/2} \mbox{Hz,}\quad \mbox{for wind}\\
  \end{array} \right.
\end{equation}
and
\begin{equation} 
\label{eq:nuc}
\nu_c=
\left\{
  \begin{array}{l}
6.37\ (p-0.46)\  10^{13}\ e^{-1.16p}\  (1+z)^{-1/2}\ \epsilon_B^{-3/2}\ n_0^{-1}\ \\ E_{kin,52}^{-1/2}\ t_{\rm days}^{-1/2}\ (1+Y_X)^{-2}\mbox{Hz,}\quad \mbox{for ISM}\\
\\
4.40\ (3.45-p)\  10^{10}\ e^{0.45p}\ (1+z)^{-3/2}\ \epsilon_B^{-3/2}\ A_{*}^{-2}\ \\  E_{kin,52}^{1/2}\ t_{\rm days}^{1/2}\ (1+Y_X)^{-2} \mbox{Hz,}\quad \mbox{for wind}\\
  \end{array} \right.
\end{equation}
Since $\nu_c$ depends on the external density, so does $Y_X$, and also $F_X$.
Equating an observed X-ray flux to the flux given by Eq. \ref{eq:aboveC} we obtain $\epsilon_B(n)$:
\begin{equation} 
\label{eq:SSC}
\epsilon_B=
\left\{
  \begin{array}{l}
 4\! \times 10^{-5}\! f(p) \bigg( \!\frac{\epsilon_e}{0.1}\!\bigg)^{2p^2\!-\!14p\!+\!12 \over p^2\!+\!2p\!-\!16}\! F_{GeV,\!-6}^{2(12\!-\!p^2)\over p^2\!+\!2p\!-\!16} F_{X,\!-5}^{2(p\!-\!4)(p\!+\!2) \over p^2\!+\!2p\!-\!16}\\ n_0^{(2\!-\!p)(p\!+\!2) \over p^2\!+\!2p\!-\!16}\! 
 t_{GeV,-3}^{(3p\!-\!2)(12\!-\!p^2) \over 2(p^2\!+\!2p\!-\!16)}\! t_X^{(3p^2\!-\!12p\!+\!4)(p\!+\!2) \over 2(p^2\!+\!2p\!-\!16)}\! d_{L28}^{8(p\!-\!2) \over p^2\!+\!2p\!-\!16}\mbox{ for ISM}\\
 \\
 2\! \times \! 10^{-4}\ \tilde{f}(p)\ \bigg( \! \frac{\epsilon_e}{0.1}\! \bigg)^{p-2 \over 3-p} F_{GeV,-6}^{p-4 \over 3-p} F_{X,-5}^{4-p \over 3-p} \\ A_{*}^{p-2\over 3-p}\  t_{GeV,-3}^{p-4 \over 4(3-p)}\ t_X^{12-5p \over4(3-p)}\ \mbox{ for Wind}\\
 \end{array} \right.
\end{equation}
where $t_{GeV},t_X$ are the times of the GeV and the X-ray observations in days, and $f(p), \tilde{f}(p)$ are dimensionless functions such that $f(p)=\tilde{f}(p)=1$ for $p=2.5$
and $f(p)=10, \tilde{f}(p)=15$ for $p=2.1$. 
All the pre-factors are chosen such that they are expected to be close to 1, and therefore the leading numerical values are an order of magnitude estimate for $\epsilon_B$ in this regime.
It is evident that very low values of $\epsilon_B$ are required in order for the X-ray and the GeV observations to be compatible with each other in the SSC suppressed type solution. This is expected, as very weak magnetic fields are essential in order to have a sufficiently strong SSC cooling.

\end{subsection}

\begin{subsection}{Slow cooling case}
\label{S:slowcooling}
We assume now that $\nu_m<\nu_X<\nu_c<\nu_{GeV}$. In this case the X-ray flux is given by \citep{Granot(2002)}:
\begin{equation} 
\label{eq:belowC}
F_\nu=
\left\{
  \begin{array}{l l}
0.461\ (p-0.04)\ e^{2.53p}\ (1+z)^{3+p\over 4}\ \bar\epsilon_{e}^{\, \, p-1}\ \epsilon_B^{1+p\over4} \\ \ n_0^{1/2}\ E_{kin,52}^{3+p\over 4}\  t_{\rm days}^{3(1-p)\over4}\ d_{L28}^{-2}
\ (\frac{\nu}{\nu_{14}})^{1-p\over 2} \mbox{mJy,}\quad \mbox{for ISM}\\
\\
 3.82\ (p-0.18)\ e^{2.54p}\ (1+z)^{5+p\over 4}\ \bar\epsilon_{e}^{\, \, p-1}\ \epsilon_B^{1+p\over 4}\\ \ A_{*}\ E_{kin,52}^{1+p\over4}\  t_{\rm days}^{1-3p\over 4}\ d_{L28}^{-2}
\ (\frac{\nu}{\nu_{14}})^{1-p\over 2} \mbox{mJy,}\quad \mbox{for wind}\\
  \end{array} \right.
\end{equation}
Comparing the observed X-ray flux at late times  with the flux given by Eq. \ref{eq:belowC} we obtain $\epsilon_B(n)$ or $\epsilon_B(A_*)$:
\begin{equation} 
\label{eq:slow}
\epsilon_B=
\left\{
  \begin{array}{l}
 10^{-5}\ g(p)\ \bigg(\frac{\epsilon_e}{0.1}\bigg)^{2(p-1) \over (P+4)} F_{GeV,-6}^{-2(p+3) \over p+4} F_{X,-5}^{2(p+2) \over p+4}\\  n_0^{-(p+2) \over p+4}\ t_{GeV,-3}^{(2-3p)(p+3) \over 2p+8}\ t_X^{3(p+2)(p-1) \over 2p+8}\ d_{L28}^{-4 \over p+4} \mbox{ for ISM}\\
 \\
 4\times 10^{-6}\ \tilde{g}(p)\ \bigg(\frac{\epsilon_e}{0.1}\bigg)^{p^2+5p-6 \over 4}  F_{GeV,-6}^{-1} F_{X,-5}^{p+2 \over 1+p}\\ A_*^{-(p+2) \over 1+p}\ t_{GeV,-3}^{2-3p \over 4}\ t_X^{(3p-1)(2+p) \over 4(1+p)}\ d_{L28}^{-2} \mbox{ for Wind.}\\
 \end{array} \right.
\end{equation}
where $g(p), \tilde{g}(p)$ are dimensionless functions such that $g(p)=\tilde{g}(p)=1$ for $p=2.5$
and $g(p)=4, \tilde{g}(p)=0.9$ for $p=2.1$. 
Once again, $\epsilon_B$ should be very low to allow for this type of solution.

\end{subsection}

\begin{figure*}
\centering
\includegraphics[width=0.48\textwidth]{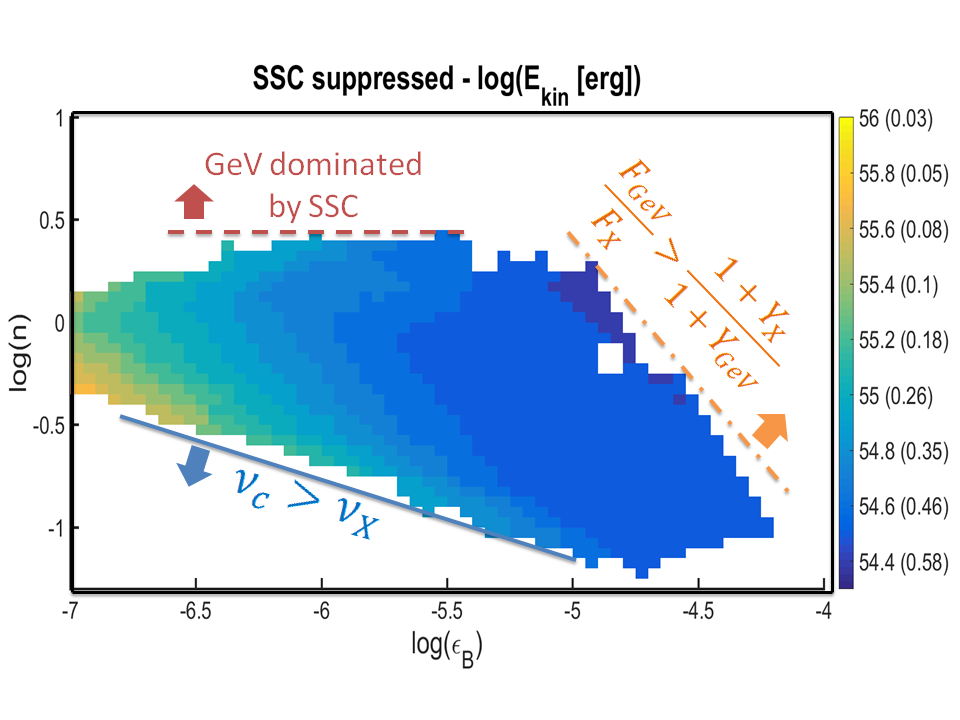}
\includegraphics[width=0.48\textwidth]{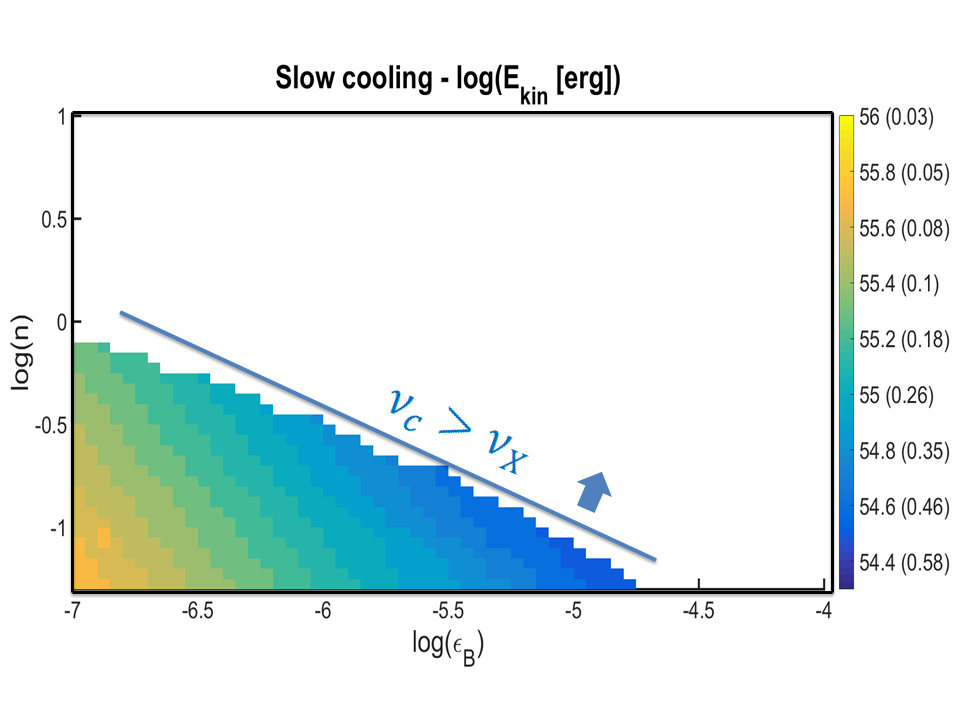}
\caption
{\small The allowed parameter space for GRB 080916C in an ISM environment. For any given point in the $\epsilon_B$-$n$ plane, colours depict the minimum possible value of the isotropic equivalent
kinetic energies for which a successful modeling is found (see the values in the colour bar). Lower limits on $E_{0,kin}$ correspond to upper limits on the prompt efficiency (in parenthesis).
The left panel corresponds to the ''SSC suppressed" solutions discussed in \S \ref{sec:model} and the right panel to the ``slow cooling" solutions discussed in the same section.
For reasonable values of $3\times10^{-2}$  cm$^{-3}$ $\lesssim n \lesssim 3$ cm$^{-3}$, we obtain: $\epsilon_{\gamma}\lesssim 0.55$, $E_{0,kin}\gtrsim 3 \times 10^{54}$ergs and $\epsilon_B \lesssim 5 \times 10^{-5}$, independent of the type of solution.}
\label{fig:080916C}
\end{figure*}

\begin{figure*}
\includegraphics[scale=0.4]{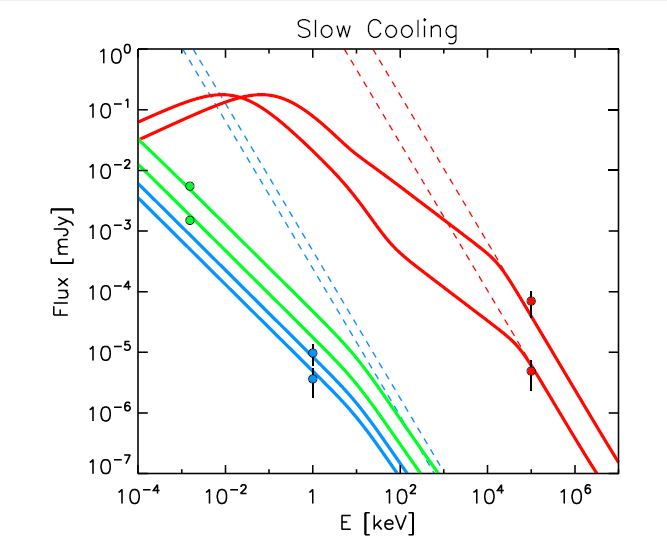}
\includegraphics[scale=0.4]{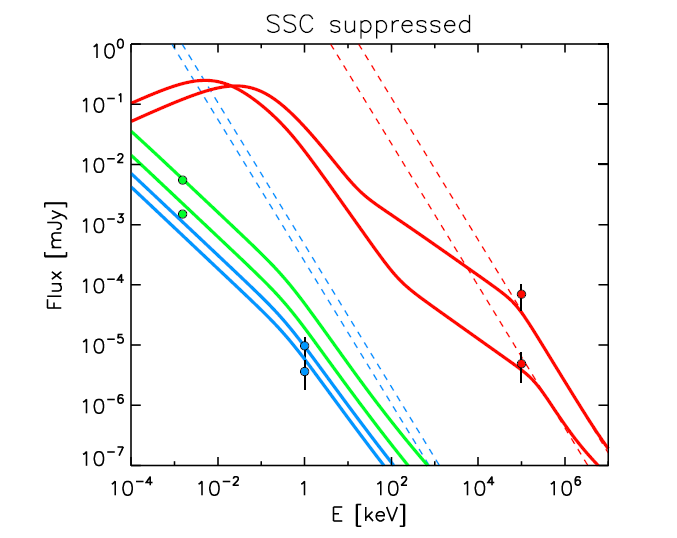}
\caption{Observed and predicted spectra for GRB 080916C in a homogeneous medium for a single model at six different times: two LAT observation times (red, 0.0046 and 0.001 days) two X-ray observation times (blue, 6.94 and 11.57 days)
and two optical observation times (green, 1.39 and 3.47 days).
Left: an example of modelling where the X-ray band is dominated by synchrotron radiation and resides below the cooling frequency, while the GeV band 
is also dominated by synchrotron radiation but resides above the cooling frequency. This situation corresponds to a typical ``slow cooling'' case. The results are shown for $\epsilon_B=4.5\times10^{-7}, n=0.14\mbox{ cm}^{-3}, E_{kin}=1.5\times 10^{55} \mbox{erg}, p=2.43$.
Right: an example of modelling where X-rays are produced by IC-suppressed fast cooling synchrotron emission, while the GeV radiation is produced by non-suppressed fast cooling synchrotron emission (corresponding to a typical ``SSC suppressed'' case).
Results are shown for $\epsilon_B=3.5\times10^{-7}, n=0.63\mbox{ cm}^{-3}, E_{kin}=1.2 \times10^{55} \mbox{erg}, p=2.36$. 
The solid curves are the results of the numerical calculation at different observation times: LAT in red, X-rays in blue and optical in green.
The observations are denoted by filled circles with error-bars (the optical error-bars are smaller than the size of the data points).
The red and blue dashed lines are the expected synchrotron spectrum above $\nu_c$, at the time of LAT and XRT observations, under the assumption the flux is not suppressed by IC (Eq. \ref{eq:aboveC}).
It is evident that, according to this modelling, this assumption is correct for the GeV data, while it is not correct for X-ray observations, either because the X-ray observations lie below $\nu_c$ (left panel) or because at the X-ray frequency the flux
is strongly suppressed by IC (right panel).}
\label{fig:SED}
\end{figure*}

\section{Numerical Modeling}
\label{sect:results}
The analytic estimates show that for reasonable external density parameters
($n \approx 1\,\mbox{cm}^{-3}$ or $A_* \approx 1$), $\epsilon_B$ should be very low in both the SSC suppressed and the slow cooling scenarios.
Combining the estimates in \S \ref{S:SSC}, \S \ref{S:slowcooling} we obtain: $10^{-6}\lesssim \epsilon_B \lesssim 10^{-3}$.
This result is in agreement with recent findings \citep{KBD(2009),KBD(2010),RBD(2011),RBD(2014),Santana(2014),Zhang(2014),Wang(2015)}.

We turn now to a detailed numerical modeling of the afterglow, which might also help us to distinguish between the two solutions.
We numerically model the afterglow emission and compare the expected fluxes with the GeV, X-ray and optical observations listed in Table \ref{tbl:data}.
In the numerical model we relax all the simplifying assumptions made above. In particular we relax the assumptions that $\nu_c$ is below the GeV band at $t_{GeV}$ or that the GeV and X-ray fluxes are dominated by synchrotron (and not SSC).

We calculate the synchrotron and SSC spectra following \cite{Granot(2002)} and \cite{Nakar(2009)}. In particular, the method developed by \cite{Nakar(2009)} allows us to compute the Compton parameter $Y(\nu)$ as a function of the electron energy and
to include possible corrections due to KN, that may affect the shape of the synchrotron spectrum.
We ignore the fact that at a given moment in time we observe emission from different emitting radii \citep{Granot(1999)}. We used \citep{Granot(1999),Granot(2002)}, to verify that the magnitude of this effect cannot change the flux as compared with the spectra we are using by more than $\sim50\%$.
For each burst, we consider values of $\epsilon_B, n$ and $p$ in the range $10^{-8}<\epsilon_B<10^{-2}$, $2.1<p<2.8$, and $10^{-2}$ cm$^{-3}$ $<n<10^2$ cm$^{-3}$.
We keep the fraction of shocked energy in the electrons to $\epsilon_e \approx 0.1$.
We calculate the fluxes at $t_{GeV}, t_X$, and $t_{opt}$
(the latter is the time of optical observations, in case such observations exist) at the observed frequency ($100\,$MeV, $1\,$keV, $\sim$eV respectively). 
We consider a model as acceptable if the difference between the estimated fluxes and the data points is within the errors 
(where the errors take into account both the uncertainties of the observations and of the model).
We remark that for some bursts it is possible that more detailed observations (i.e. in more frequency bands and at different times or using the spectral indices in the available observed bands) will further constrain the allowed parameter space \footnote{In general, errors on spectral indices can be large.
Instead of using spectral indices, we consider (both for X-rays and optical) two data points at times as far as possible one from
each other (see \S \ref{simultaneous} where we do use spectral indices). This assures that we are recovering the correct slope of the light-curve.
For the GeV we consider two data points after $T_{90}$, in order to avoid possible contamination from the prompt. 
We are therefore considering the data point where the flux is smaller, and the errors on the spectral index and temporal decay can be very large.
However, we know that LAT spectra that are rising in $\nu F_{\nu}$ are rarely observed and in any case their photon index is never harder than -1.7. 
In order to consider only realistic solutions, we add this requirement on the LAT spectral index.}.
However, since we are mainly interested in limits, this only means that at worst our results may be ``too conservative".

Optical observations are available for eight out of ten bursts in our sample
and we include them in the modeling.
The optical data is missing for GRB 100414A and GRB 110625A. 
The X-ray band may either be above or below cooling (corresponding to the ``SSC suppressed " and ``slow cooling" cases respectively). However, there are generally no
reasonable solutions with the optical band being above the cooling frequency at times of order $\sim1\,$day, with the possible exception of GRB~090926A
for which such a solution is only marginally consistent for low values of $p \approx 2.1$ (see also \citealt{Cenko(2011)}).
For all considered GRBs we can successfully reproduce the observed fluxes within the general synchrotron or synchrotron self Compton model described above.
This supports the interpretation of GeV photons as radiation from the external forward shock, at least for those bursts considered in this study, which are all the bursts with a long lasting, power law decaying GeV emission.

We present, first, detailed results for one GRB and then we discuss more generally the results obtained for all bursts in our sample. 
We choose to discuss the case of GRB~080916C in an ISM environment. 
For this burst both the ``suppressed SSC" and the ``slow cooling" solutions can account for the observations. 
In addition, there are some solutions at large densities ($n>30 \mbox{ cm}^{-3}$) and very weak magnetic fields ($\epsilon_B<10^{-6}$) in which both the GeV and X-ray fluxes are dominated by SSC instead of synchrotron emission.
However, a more careful examination reveals that all the SSC dominated solutions for this burst correspond to a GeV photon index $\Gamma>-1.7$ ($N_\nu\propto\nu^{\Gamma}$) which is not in agreement with observations
(for all the bursts in our sample, the photon index is known and it is never harder than -1.7. In fact, this is true even when considering the entire sample of GRBs detected by GeV, as can be seen in Fig. 25 of \citealt{ACK13}).
Fig. \ref{fig:080916C} depicts the allowed parameter space separately for the ``SSC suppressed " (left panel) and the ``slow cooling" (right panel) solutions. 
The colored region depicts the allowed region in the $\epsilon_B$-$n$ plane. 
In some cases, for a given pair of values ($\epsilon_B, n$) it is possible that several solutions
are found, corresponding to different vales of $E_{0,kin}$ and/or $p$.
In this case we show the lowest value of $E_{0,kin}$, regardless of the value of $p$. The reported values of $E_{0,kin}$ should then be considered as lower limits.
The Compton parameter $Y_X$ can vary between 3-100. At the lower end of this range, $Y_X$ is affected by KN suppression and may differ somewhat from the expression derived for the Thomson regime (Eq.~\ref{Yx}) used for the analytic estimates presented in \S \ref{sec:model}. 
Moreover, $Y_{GeV}$ can vary between 0.1-3, implying that a moderate suppression of the GeV flux might also take place, but only when $\epsilon_B$ is very small and $n$ is large.
For $Y_{GeV}\lesssim0.3$ and for X-rays in Thomson regime, the numerical results are indeed compatible with the analytical formulas presented in Eqns. \ref{eq:SSC}, \ref{eq:slow}.
The shape of the allowed parameter space can be understood in the following way. First, the border to the lower left of the ``SSC suppressed " region is defined by $\nu_X=\nu_c$ at the time of X-ray observations. Below this line, the X-ray flux is produced by slow cooling electrons
(this roughly corresponds to the upper right border of the ``slow cooling" solutions, where some superposition is allowed due to variations of the parameters $E_{0,kin}$ and $p$). The upper border in both cases is due to the fact that for larger densities the GeV component becomes dominated by SSC instead of synchrotron radiation.
Finally the upper right border for the ``SSC suppressed" solutions arises from the requirement that $Y_X$ should be large enough as compared with $Y_{GeV}$ in order to account for the flux discrepancy between the GeV and the X-rays.
All the allowed solutions correspond to isotropic equivalent kinetic energies satisfying $E_{0,kin}\gtrsim 3\times 10^{54}\,$ergs, and imply that the prompt efficiency is moderate: $\epsilon_{\gamma}\lesssim 0.55$. This limit on the efficiency rules out the value inferred just from the X-ray observations ($\epsilon_{\gamma}=0.85$) that is derived assuming that X-rays are above $\nu_c$.
The allowed parameter space results in strong upper limits on the magnetization: $\epsilon_B \lesssim 5 \times 10^{-5}$, regardless of the type of solution. Strong limits on the magnetization ($\epsilon_B\lesssim 3\times 10^{-4}$) can be obtained also for solutions in a wind medium, consistent with previous estimates of the magnetization for this burst \citep{KBD(2009),Gao(2009),Zou(2009),Feng(2010)}.
Typical SEDs (spectral energy distributions) for the ``SSC suppressed" and ``slow cooling" cases for this burst (for a particular set of parameters) are shown in Fig. \ref{fig:SED}. The solid curves are the SEDs resulting from numerical calculation at six different times: at the time of the two LAT observations (red), at the time of the two X-ray observations (blue), and at the time of the two optical observations (green). 
The six observations are denoted by filled circles with error-bars (the optical error-bars are smaller than the size of the data points).
The red and blue dashed lines are the expected synchrotron spectrum above $\nu_c$, at the time of LAT and XRT observations, under the assumption that the flux is not suppressed by IC (Eq. \ref{eq:aboveC}).
It is evident that, according to this modelling, this assumption is correct for the GeV data, while it is not correct for X-ray observations, either because the X-ray observations lie below $\nu_c$ (left panel) or because at the X-ray frequency the flux
is strongly suppressed by IC (right panel).

A similar analysis has been performed for all bursts in our sample, for both constant and wind-like density profiles of the external medium.
For a constant profile (ISM), four bursts have both ``SSC suppressed" and ``slow cooling" type solutions, while one has only a ``SSC suppressed"
solution and another has only a ``slow cooling" solution. For the remaining bursts (090510, 090902B, 100414, 110625) it is impossible to find acceptable models to the data, for an ISM. For a wind-like medium, no bursts have ``SSC suppressed" type solutions, whereas seven bursts have ``slow cooling" solutions.
For the remaining three bursts (090926A, 090902B and 090510) a good modeling of the data is possible, but it does not correspond to any of the solutions discussed so far. In fact, for these bursts the GeV flux is dominated by SSC (but still
consistent with the requirement on the GeV spectral slope: $\Gamma<-1.7$).
whereas the X-ray flux is dominated by synchrotron from fast cooling electrons (for GRB 090926A X-rays may also be dominated by SSC in the solution for a wind type medium)\footnote{It is possible that a pure synchrotron solution may be found also for GRB 090510 and 090902B; however, it would require some changes of the assumptions of the present model,
such as a more general power-law profile of the external density, a deviation from a single power-law distribution of electrons, etc.  Indeed for GRB 090510, previous studies proposed synchrotron solutions, but these relied on either very low external densities, energy injection or evolution of the micro-physical parameters
\citep{De Pasquale(2010),Corsi(2010)}. For GRB 090902B, \cite{RBD(2011)} have found synchrotron solutions but with a deviation from a single power-law distribution of electrons  (and even in this case the limits on $\epsilon_B$ remain very strong).}.

\begin{figure*}
\centering
\includegraphics[width=0.45\textwidth]{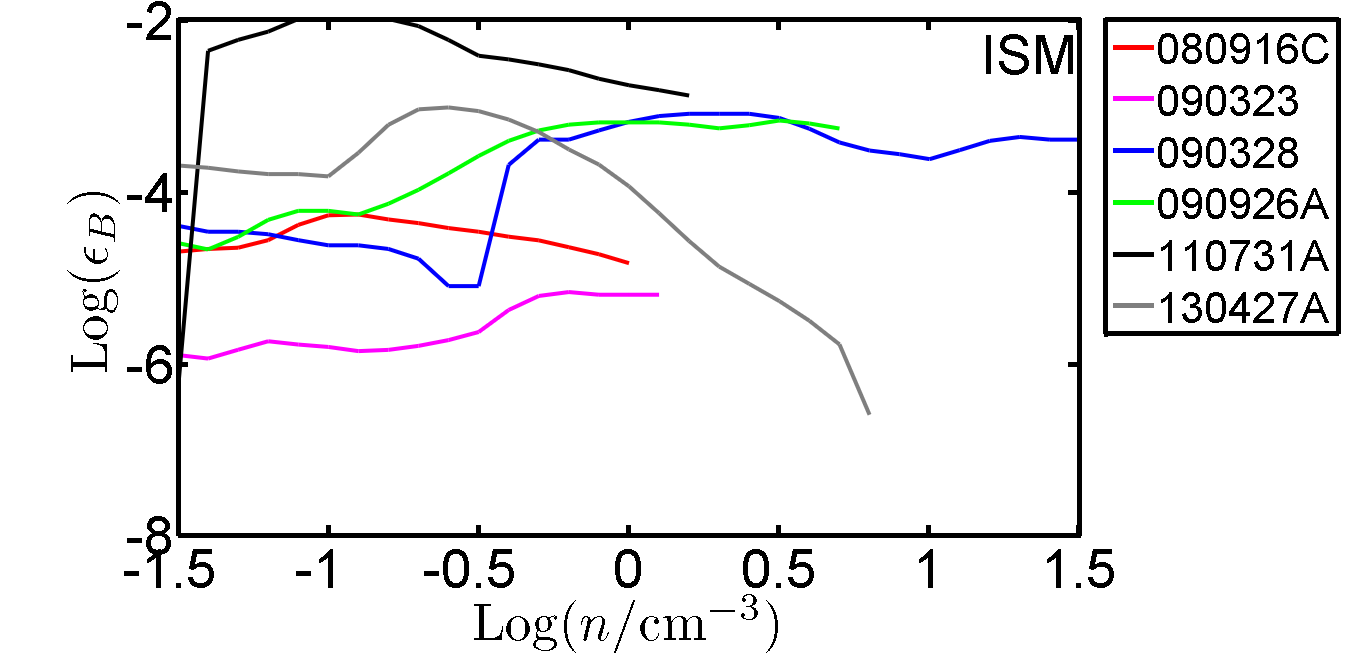}
\includegraphics[width=0.45\textwidth]{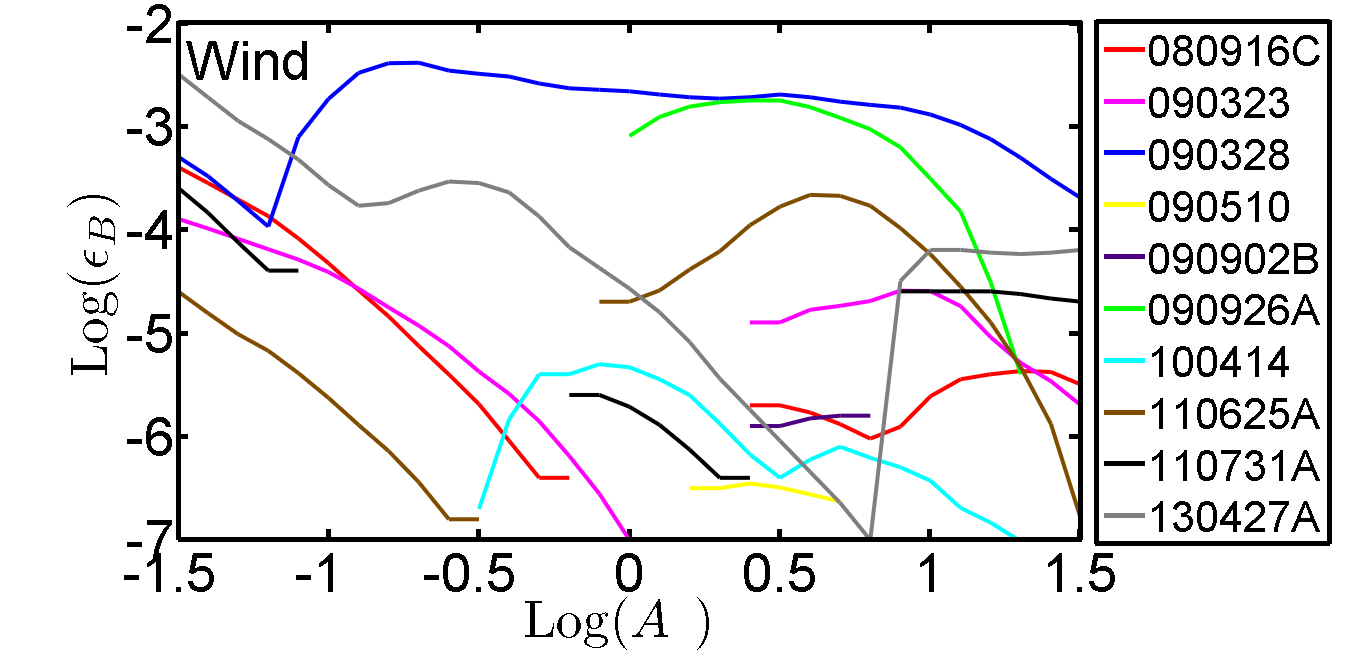}
\includegraphics[width=0.45\textwidth]{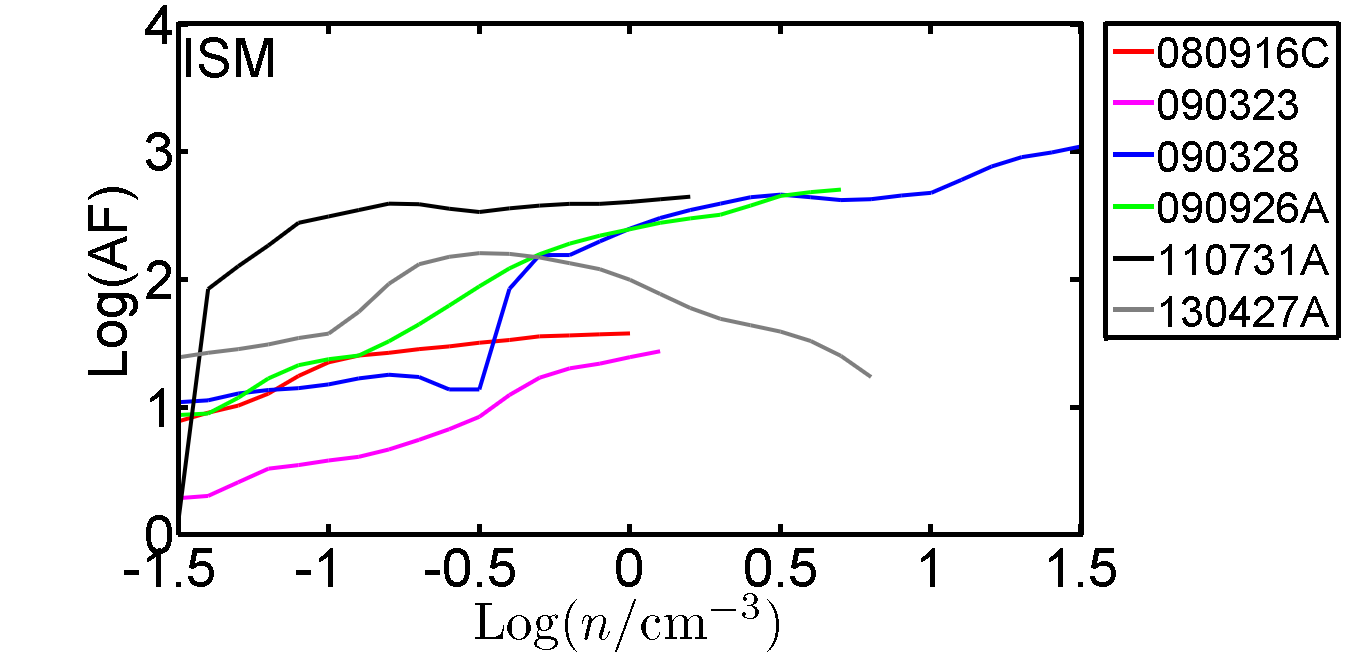}
\includegraphics[width=0.45\textwidth]{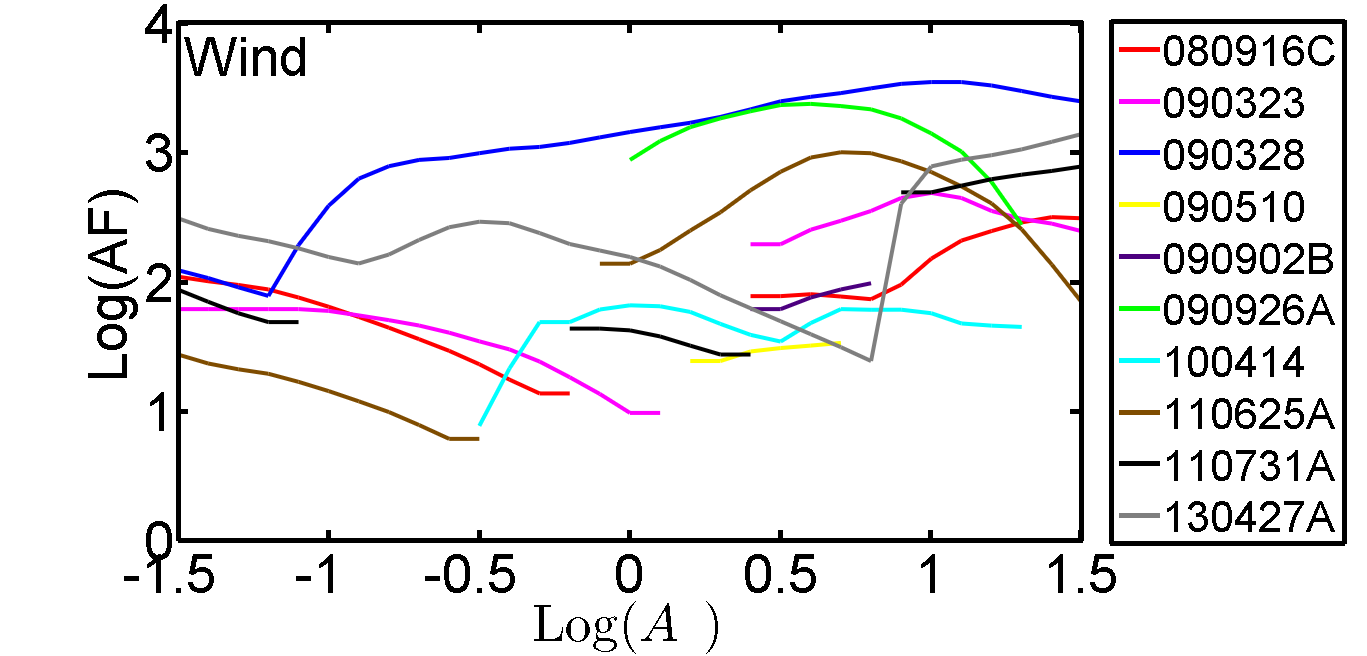}
\caption
{\small Top panels: upper limits on $\epsilon_B$ as a function of the external density for ISM (top left) and wind (top right) environments.
Lower panels: upper limits on the logarithm of the magnetic field amplification factors (AF) beyond shock compression (for ISM in the lower left panel and wind in the lower right panel), assuming a seed (unshocked) magnetic field of $10\,\mu$G.}
\label{fig:epsBn}
\end{figure*}

\begin{figure*}
\centering
\includegraphics[width=0.45\textwidth]{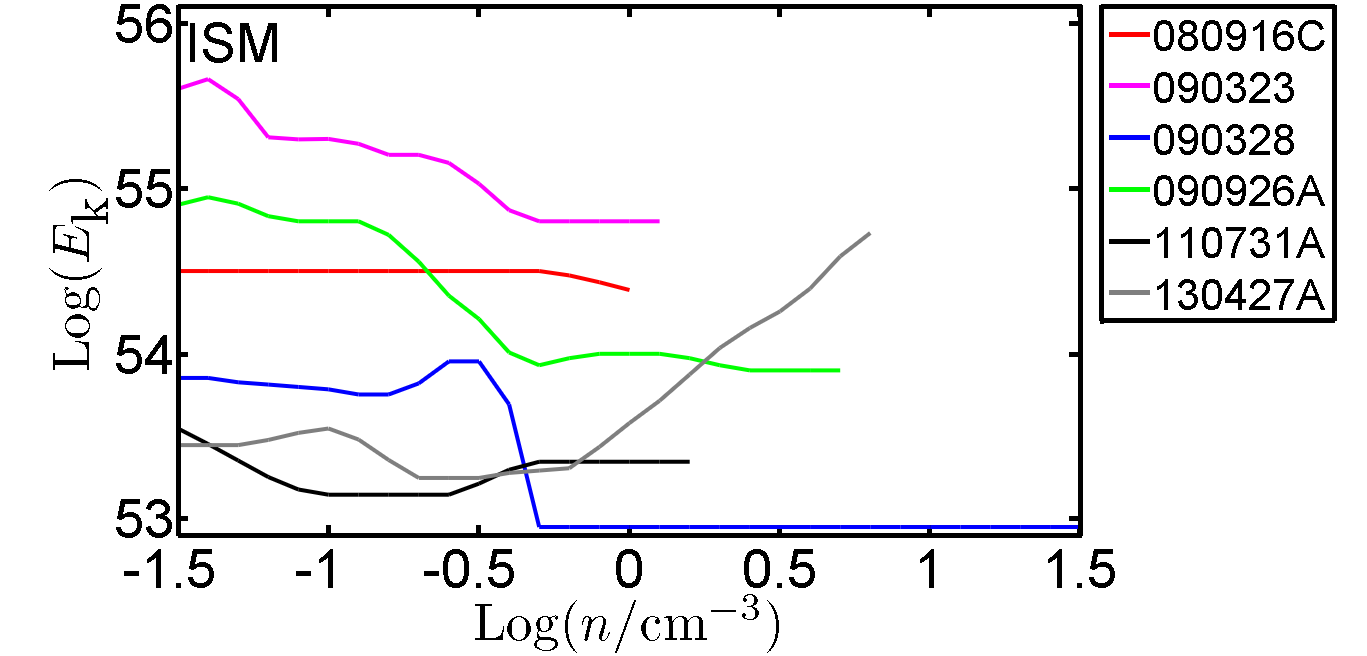}
\includegraphics[width=0.45\textwidth]{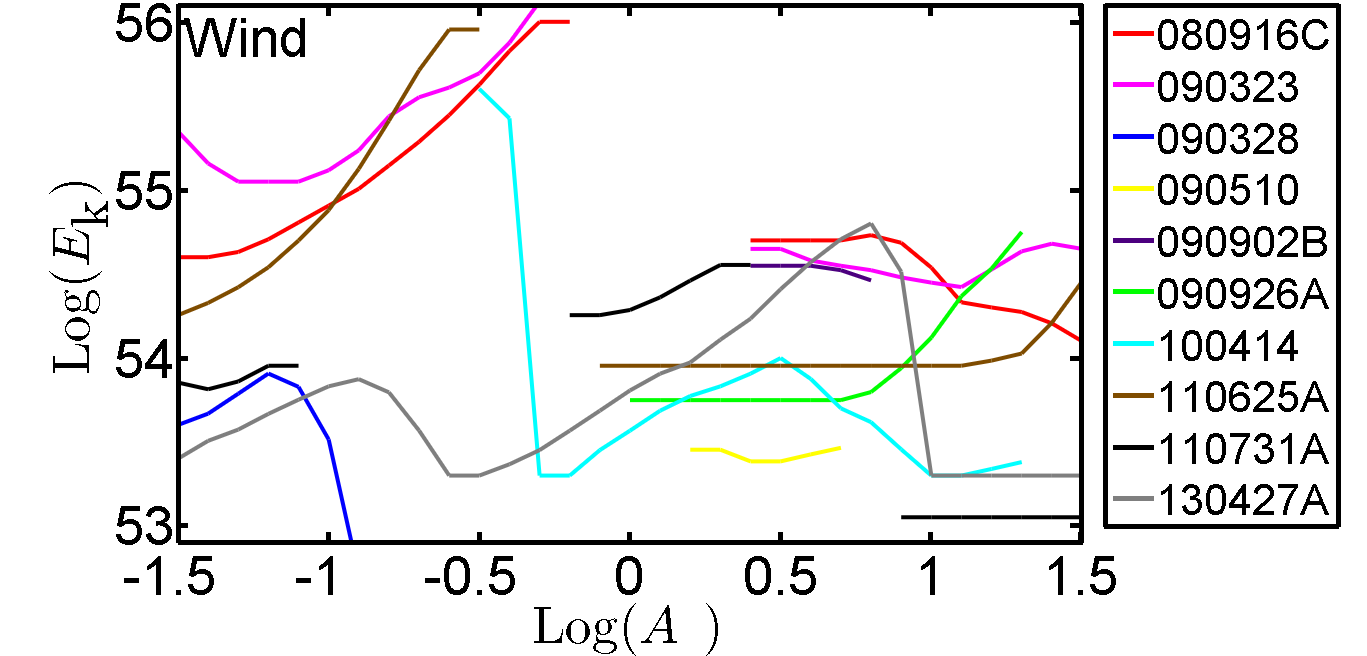}
\caption
{\small Lower limits on the isotropic equivalent kinetic energies as a function of the external density for ISM (left panel) and wind (right panel) environments.
The limits range between $10^{53}-3\times 10^{55}$erg and are generally much larger than the kinetic energies estimated from X-rays only, assuming fast cooling synchrotron.}
\label{fig:Elimit}
\end{figure*}

\begin{figure*}
\centering
\includegraphics[width=0.45\textwidth]{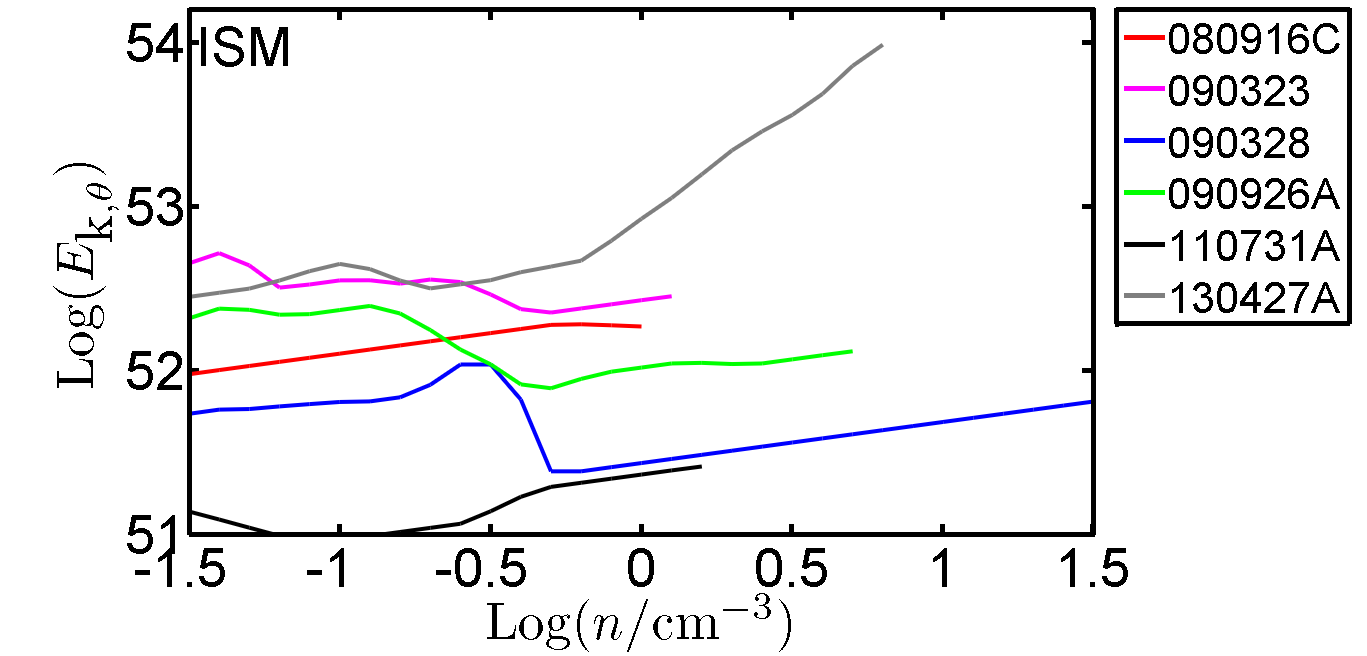}
\includegraphics[width=0.45\textwidth]{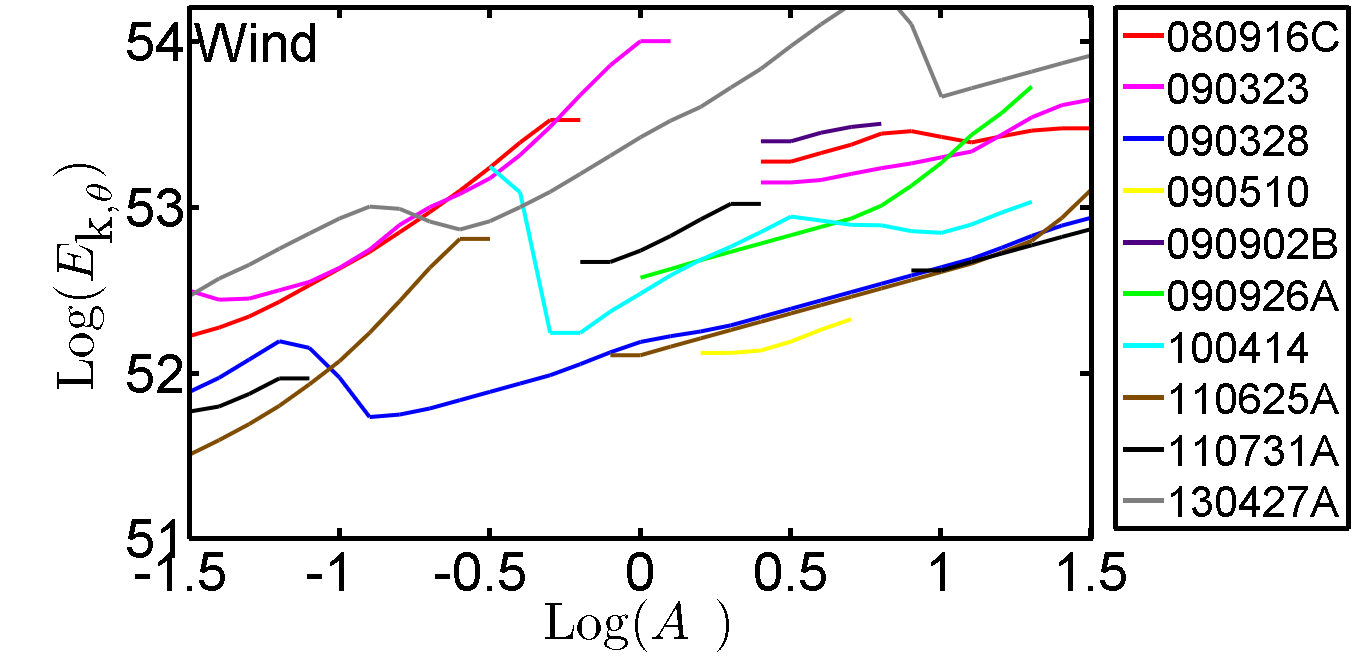}
\caption
{\small Lower limits on the true (collimated) kinetic energies as a function of the external density for ISM (left panel) and wind-like (right panel) environments.
All GRBs are consistent with an energy $\lesssim 10^{52.5}\,$erg.}
\label{fig:Etrue}
\end{figure*}

Fig. \ref{fig:epsBn} depicts the upper limits on $\epsilon_B$ as a function of the external density for each burst, for an ISM (left upper panel) and for a wind (right upper panel). The bottom panels show the required amplification factor of the magnetic field beyond shock compression,
denoted $AF$ by \cite{Santana(2014)}, for the given $\epsilon_B$ and $n$, for a seed field of $10\,\mu$G.
The upper limits are approximately $10^{-5}<\epsilon_B<10^{-3}$ and $3<AF<1000$ for both ISM and wind environments.
Lower limits on the (isotropic equivalent) kinetic energies as a function of the external density for both ISM and wind are shown in Fig. \ref{fig:Elimit}.
The limits range between $10^{53}-3\times 10^{55}\,$erg and are generally
much larger than the kinetic energies estimated from X-rays only, assuming fast cooling synchrotron \footnote{The only case where the kinetic energy is not much larger than that estimated from the X-rays is that of GRB 090328, for an ISM and large densities $n \geq 1\mbox{ cm}^{-3}$. In this case the solutions are such that the GeV is
dominated by SSC instead of synchrotron and the kinetic energy is similar to that estimated from the X-rays assuming fast cooling synchrotron.}.
These limits on the isotropic energy can be translated into limits on the collimation corrected energy if the jet opening angle is known (or if a lower limit on the jet opening angle can be estimated from the non-detection of
a steepening in the afterglow light curve - i.e., a jet break). The measurements/limits on the jet break times are listed in Table \ref{tbl:data} (one before last column).
These have been translated into estimates/limits on the jet opening angles \citep{Sari(1999),Chevalier(2000),Frail(2001)}.
Fig. \ref{fig:Etrue} shows the derived limits on the true (collimated) jet kinetic energy as a function of density. 
All GRBs are consistent with an energy $\lesssim 10^{52.5}\,$erg.
The shapes of the curves in Figs. \ref{fig:epsBn}, \ref{fig:Elimit}, \ref{fig:Etrue} are quite complex and can vary from burst to burst. This is due to the many different parameters involved in determining these limits.
For low enough densities ($0.01 \mbox{ cm}^{-3} \lesssim n \lesssim 1\mbox{ cm}^{-3}$, $0.01 \lesssim A_* \lesssim 1$), the solutions are usually of the ``slow cooling" type.
In this regime, the upper limit on $\epsilon_B$ is determined by the condition $\nu_c=\nu_X$. In some bursts (but for ISM only), for somewhat larger densities ($0.1 \mbox{ cm}^{-3} \lesssim n \lesssim 3\mbox{ cm}^{-3}$), there are ``SSC suppressed" type solutions, and in this case the limits on $\epsilon_B$ arise
from the requirement that $Y_X$ should be sufficiently large to account for the ratio of fluxes between GeV and X-rays.
For larger densities ($1 \mbox{ cm}^{-3} \lesssim n \lesssim 10\mbox{ cm}^{-3}$, $1 \lesssim A_*  \lesssim 10$) there might be solutions where the GeV flux becomes dominated by SSC, and, if such a solutions exist they are limited by the fact that as $\epsilon_B$ increases the SSC peak becomes too weak and
the flux in the GeV band is no longer sufficiently large as compared with that in X-rays. Finally, for the largest densities ($10 \mbox{ cm}^{-3} \lesssim n \lesssim 100\mbox{ cm}^{-3}$, $10 \lesssim A_*  \lesssim 100$)
the flux may become dominated by SSC both in the GeV band and in the X-rays. Since the optical band in this case is still dominated by synchrotron, the allowed solutions are then limited by the fact that as $\epsilon_B$ increases the SSC peak becomes weaker and cannot account simultaneously for the GeV and optical fluxes.

In principle, another constraint on these solutions arises from the fact that there is no evidence for a passage of the cooling break through the X-ray band during the time of
observation. This constraint is only relevant for bursts in a wind environment (for which $\nu_c$ increases with time) and ``SSC suppressed " solutions, 
and for bursts in an ISM environment (where $\nu_c$ decreases with time) for the ``slow cooling" solutions.
However, due to the fact that X-ray observations are typically available for less than 2 decades in time (say 0.1-10 days), and since $\nu_c$ evolves only slowly with time
($\nu_c \propto t^{-1/2}$ for an ISM environment and $\nu_c \propto t^{1/2}$ in a wind), $\nu_c$ cannot change by more than an order of magnitude during the duration of the observation.
This is less than the spectral range of XRT (0.3-10 keV), and since we are using integrated fluxes in this work, it will result in an undetectable spectral change
given the quality of the data.


\subsection{Results from modelling of simultaneous X-ray and GeV data}
\label{simultaneous}

For four bursts in our sample, early time X-ray observations (simultaneous to observations in the GeV band) are available. So far we have considered only late time X-ray data, because our aim is to test if these data can be safely used to infer the blast wave kinetic energy. 
In this section we exploit early time simultaneous X-ray/GeV detections (when available) in order to test the consistency of our modelling and possibly reduce the parameter space identified by the modeling of late time X-ray data.

For all bursts we find that the parameter space overlaps with the one inferred from late time observations. In other words, it exists a region of the parameter space that allows a good description of both early and late time X-ray and GeV fluxes. In the following we discuss in detail our findings, for each one of these bursts.

\subsubsection{090510}
For this burst we use data at 114 seconds, taken from \cite{De Pasquale(2010)}. Similarly to when we use late time X-ray data, we do not find a good match between the observations and the numerical calculations for a homogeneous medium.
For a wind medium we find values of $\epsilon_B=10^{-7}-3\times 10^{-5}$, $E_{0,kin}=8\times10^{52}-6\times 10^{53}$ergs, $A_*=3-30$. These ranges of values are similar to those obtained with the late time data, but somewhat wider. These parameters correspond to a situation in which the X-rays are either dominated by fast cooling synchrotron or by SSC (for the larger densities) emission, whereas the GeV band is always dominated by the SSC component.
An example of modelling (where the X-ray flux is dominated by synchrotron) is shown in Fig. \ref{fig:sim} (upper left panel).
The dashed lines show the separate contributions from synchrotron and SSC, and the solid line is their sum.
X-ray and GeV observations, as well as the errors in the fluxes and spectral slopes are denoted by the blue and red error-bars correspondingly.
The dotted line is an extrapolation of the expected synchrotron spectrum above $\nu_c$ (Eq. \ref{eq:aboveC}), showing that in all cases, the X-rays are incompatible with lying on the same segment of the spectra as the GeV photons.
These results are similar to those obtained by \cite{Panaitescu(2011)}, who show that the 100 s data is best modelled by a fast cooling synchrotron component for the X-rays and an SSC component for the GeV.
They also find that large isotropic energies, large densities and very low $\epsilon_B$ are required. 
Other studies \citep{De Pasquale(2010),Corsi(2010)} have shown that the observations can be explained in the context of a homogeneous medium if a very small density is invoked ($n\approx 10^{-6}\mbox{ cm}^{-3}$), well below the range of densities considered in this work. These studies also outline that some deviation from the standard forward shock scenario (energy injection, evolution of the microphysical parameters in time or a structured jet model) is required  in order to obtain a good description of the observations.

\subsubsection{110625A}
For this burst we model the SED at 260 seconds. Both the X-ray and GeV data are taken from \cite{Tam(2012)}. Once again, as for the late time data, there is no available parameter space if a homogeneous medium is assumed.
For a wind medium, however, two sets of parameter ranges are found, similar to what was found from the modelling of the late time data. The first one ($\epsilon_B= 10^{-7}-6\times 10^{-5}$, $E_{0,kin}=3\times10^{53}-3\times 10^{54}\,$ergs, $A_*=0.6-30$)
corresponds to a X-rays either dominated by fast cooling synchrotron or by SSC (for the larger densities) and GeV emission always dominated by the SSC component.
The second set of parameters is found at lower densities ($A_*<0.03$), larger isotropic energies ($E_{0,kin}=10^{54}-2\times 10^{55}\,$ergs) and similar values of $\epsilon_B$. In this case, the X-rays are dominated by
slow cooling synchrotron whereas the GeV range is dominated by fast cooling synchrotron. This last type of modelling was found also by \cite{Tam(2012)} who suggest that the cooling frequency at $t\sim 300\,$s should be between $10-100\,$MeV. An example of such modelling is shown in Fig.  \ref{fig:sim} (upper right panel).

\subsubsection{110731A}
For this burst we use the data at 277 seconds, taken from \cite{ACK13b}. For a homogeneous medium we find
$\epsilon_B= 2\times 10^{-6}-10^{-2}$, $E_{0,kin}=10^{53}-2\times 10^{55}\,$ergs, $n<3\mbox{ cm}^{-3}$. For these ranges of parameters, the X-rays are always dominated by synchrotron (either slow cooling for lower densities or
fast cooling and SSC suppressed for larger densities), whereas the GeV emission is dominated either by fast cooling synchrotron or by SSC (for larger densities). These solutions are similar to those found with late time data.
For a wind medium we also find allowed parameter space but with almost no overlap with the parameter space identified with the late time data. This favours a homogeneous medium.
An example of modelling where the X-ray frequency is above $\nu_c$ but the X-ray flux is suppressed by IC and where the GeV flux is produced by un-suppressed fast cooling synchrotron is shown in Fig.~\ref{fig:sim} (bottom left panel).

\subsubsection{130427A}
\begin{figure*}
\includegraphics[scale=0.5]{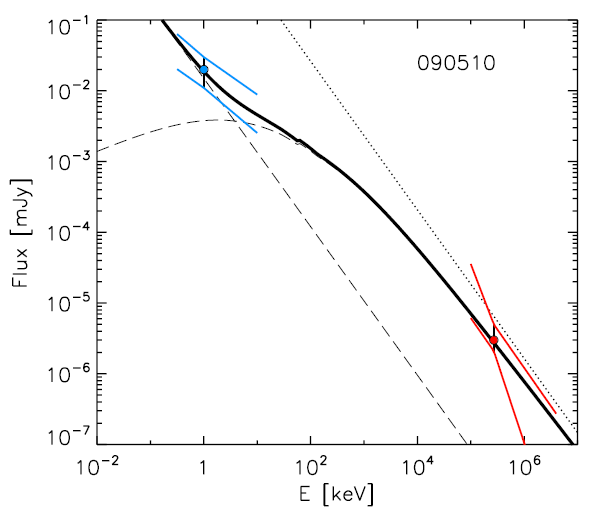}
\includegraphics[scale=0.5]{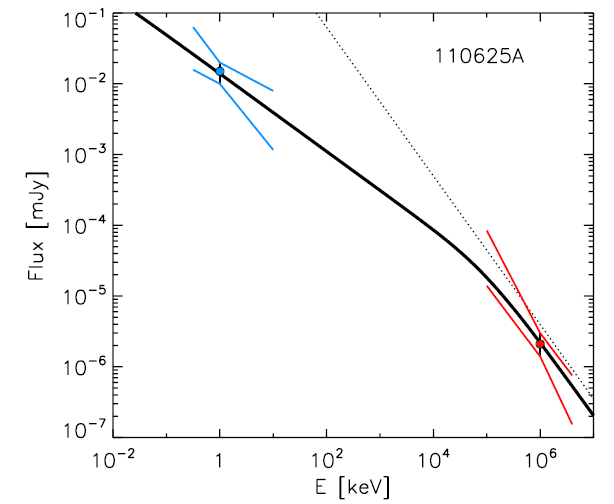}\\
\includegraphics[scale=0.5]{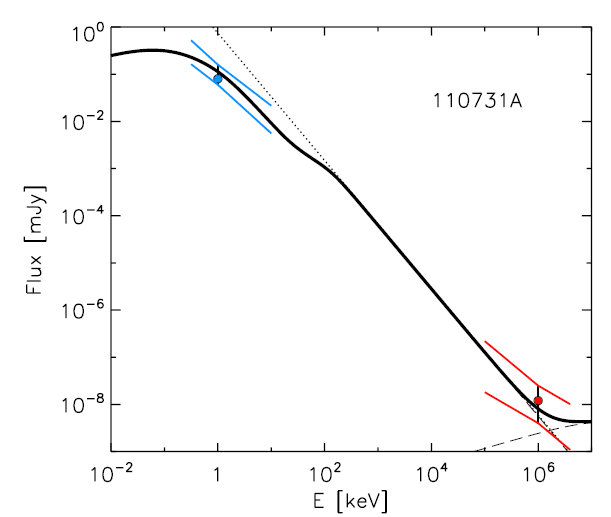}
\includegraphics[scale=0.5]{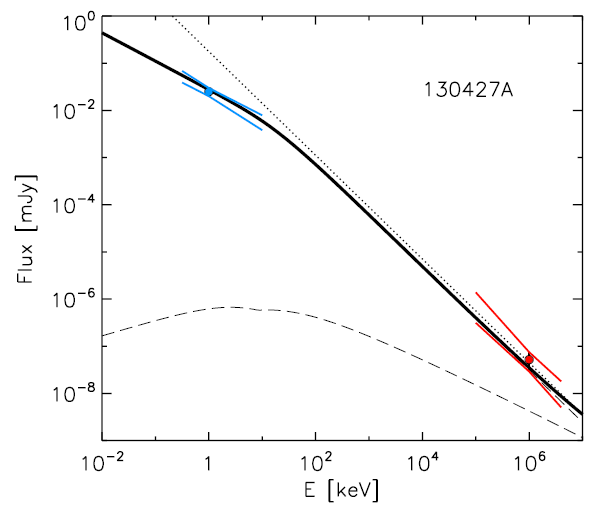}
\caption{Observed and predicted spectra for the four bursts with simultaneous X-ray and GeV data (for a particular set of parameters).
In all panels, the dashed lines show the separate contributions from synchrotron and SSC, and the solid line (which is their sum) is the total flux predicted by the model.
X-ray and GeV observations, as well as the errors in the fluxes and spectral slopes are denoted by the blue and red error-bars correspondingly.
The dotted line is the synchrotron spectrum above $\nu_c$ as predicted in case negligible IC cooling (Eq. \ref{eq:aboveC}), showing that in all cases, the X-rays do not satisfy both these requirement, being either below $\nu_c$ or suppressed by IC cooling.
Top left: SED for 090510 in a wind medium with $\epsilon_B=4\times10^{-7}, A_*=8, E_{kin}=3\times 10^{54} \mbox{ergs}, p=2.3$
(in this example X-rays are dominated by fast cooling synchrotron and the GeV is produced by SSC).
Top right: SED for 110625A in a wind medium with $\epsilon_B=1.6\times10^{-6}, A_*=0.02, E_{kin}=3.8 \times10^{54} \mbox{ergs}, p=2.1$ (X-rays are dominated by slow cooling synchrotron and the GeV by fast cooling synchrotron).
Bottom left: SED for 110731A in a homogeneous medium with $\epsilon_B=1.3\times10^{-3}, n=0.02\mbox{ cm}^{-3}, E_{kin}=7\times 10^{53} \mbox{ergs}, p=2.7$
(X-rays are dominated by SSC suppressed fast cooling synchrotron and the GeV is produced by un-suppressed fast cooling synchrotron).
Bottom right: SED for 130427A in a homogeneous medium with $\epsilon_B=2\times10^{-4}, n=0.025\mbox{ cm}^{-3}, E_{kin}=1.3\times 10^{54} \mbox{ergs}, p=2.2$
(X-rays are dominated by slow cooling synchrotron and the GeV by fast cooling synchrotron).}
\label{fig:sim}
\end{figure*}

For this burst we use the data at 25,000 seconds. The X-ray data is taken from \cite{Maselli(2014)}, and the GeV data is taken from \cite{ACK13c}. For a homogeneous medium we find
$\epsilon_B=  10^{-7}-2\times10^{-3}$, $E_{0,kin}=3\times10^{53}- 10^{56}\,$ergs, $n<2\mbox{ cm}^{-3}$. Here, the X-rays are always dominated by synchrotron (either slow cooling for lower densities or
fast cooling and SSC suppressed for larger densities), whereas the GeV emission is dominated either by fast cooling synchrotron or by SSC (for larger densities). The allowed parameter space almost completely overlaps
that obtained with the previously considered data set. For a wind medium two sets of parameters are found. The first, with $\epsilon_B=  10^{-7}-10^{-2}$, $E_{0,kin}=2\times10^{53}- 10^{56}\,$ergs, $A_*<4$.
These parameters correspond X-rays dominated by slow cooling synchrotron and GeV emission dominated by fast cooling synchrotron.
Another set of parameters is found with larger densities ($A_*>5$), smaller isotropic energies ($E_{0,kin}=10^{53}-5\times 10^{53}$ergs) and similar values of $\epsilon_B$. In this case, both the X-rays and the GeV emission are
dominated by SSC. Both these sets of parameters are found also with the previous data set presented above.
An example of modelling where the X-rays are dominated by slow cooling
synchrotron and the GeV by fast cooling synchrotron is shown in Fig.~\ref{fig:sim} (bottom right panel).

\section{Conclusions}
\label{Conclusions}

We have considered all GRBs detected both by LAT and XRT, for which the GeV emission lasts longer than the prompt and it decays as a power-law in time. We have analyzed the late time (around $3\times10^2\,$s) GeV radiation and the late time (around $1\,$day) X-ray and optical radiation (optical observations are available for 8 out of the 10 bursts in our sample).
Assuming that all these emissions are produced by electrons in the shocked circumburst medium we have estimated the conditions at this shock and have elaborated on the exact radiation process. We modeled the broad band observations as radiation from the forward external shock under two different assumptions for the external density: constant (ISM) and wind-like. We derived synchrotron and SSC spectra (accounting for the role of KN suppression) and compared the simulated SEDs with observations. While we fix $\epsilon_e= 0.1$ we have allowed all the other parameters to vary over a wide range of values.

We have found that for all bursts in our sample it is possible to account for the broad band (optical to GeV or at least X-ray to GeV) observations within the forward shock model. 
For each burst we have found a range of possible solutions, which can be classified into two families, depending on the position of $\nu_X$ as compared to $\nu_c$. In general, low densities and low $\epsilon_B$ ($n \lesssim 1 \mbox{ cm}^{-3}, \epsilon_B \lesssim 10^{-4}$) correspond to a high cooling frequency, above the X-ray band. Contrary to the electrons emitting X-rays, GeV radiating electrons are in the fast cooling regime and $\nu_X$ and $\nu_{GeV}$ do not lie in the same portion of the synchrotron spectrum.
In this case the GeV flux is a better proxy for the kinetic energy, since the X-ray flux depends not only on $E_{0,kin}$ and $\epsilon_e$ but also on $n$ and $\epsilon_B$. 
Since most of the synchrotron spectrum is produced by slow cooling electrons, these solutions require, in general, large energies.
For larger values of $\epsilon_B$ and/or larger densities the cooling frequency falls below $\nu_X$. In this case both the X-rays and GeV photons are in the high energy part of the spectrum and
in order to account for the relatively low X-ray flux as compared to the GeV flux, the X-ray flux must be suppressed via SSC cooling, while, because of the KN suppression, the GeV flux is not similarly suppressed. Also in these solutions, the GeV flux is a better proxy for the kinetic energy as compared to the inverse Compton suppressed X-ray flux, whose value depends on the Compton parameter and hence indirectly on density and $\epsilon_B$.

Even if the modeling within the forward shock model is successful, only limits to the values of the unknown parameters can be inferred, due to the degeneracy among these parameters. 
Wind solutions exist for all the explored range of densities ($10^{-2} < A_\star < 10^2$).
On the other hand ISM requires $n < 5\rm\,cm^{-3}$.
For $n$ and $A_\star$ in the range 0.1-10, the largest possible values for $\epsilon_B$ vary in the range $10^{-3}-10^{-5}$,
and the upper limits on the amplification factors of the magnetic field beyond shock compression are $3<AF<1000$.
The lower limits derived on $E_{0,kin}$ are always larger than the estimates derived using the standard assumption that the X-ray flux around one day is above $\nu_c$ and that it is not affected by Inverse Compton cooling. 
The inferred values of the kinetic energy are sometimes large, but they reflect the 
isotropic equivalent energy, and not the true energy of the blast wave. Moreover, we are considering here a sample of very energetic bursts, among the most energetic ever detected. Their energetics are not representative of the whole sample, since they likely belong to the high-energy tail of the population.
The lower limits on $E_{0,kin}$ translate into upper limits on the efficiency of the prompt mechanism $\epsilon_\gamma$. These upper limits
are always less than $50\%$.

The values of the magnetic field inferred in our analysis are only slightly above the values that would arise from a simple shock compression of the magnetic field.
It is interesting to note that
our results are in line with a recent proposal by \cite{Lemoine(2013)}. In this work \cite{Lemoine(2013)} suggest
that shocks amplify the magnetic field significantly (this is required for particle acceleration within these shocks). However, the downstream magnetic field decreases rapidly with the distance from the shock. In this case the optical and X-ray observations probe values of the magnetic field far from the shock.
The GeV emitting electrons cool rapidly in a region closer to the shock, where the magnetic field is still large. However, as the GeV, fast cooling, flux is almost independent of the magnetic field,
the value of $\epsilon_B$ used to model GeV radiation does not affect too much the results. A good modeling can be found by using the same, small $\epsilon_B$ for optical, X-ray and GeV observations, with minor corrections to the GeV flux in the scenario outlined by \cite{Lemoine(2013)}. 

The sample considered in this work includes only bursts with temporally extended emission detected by LAT.
It is unclear whether or not these bursts are intrinsically different from bursts without GeV emission and represent a separated population. 
However, we note that the bursts detected in the GeV range are generally the brightest ones in the sub-MeV range \citep{Swenson(2010),Beniamini(2011),ACK13}.
This supports the possibility that this kind of emission is typically present in GRBs, but it can be detected only for the most fluent bursts, being below the LAT detection threshold for fainter GRBs.
If this is the case, then the bursts studied in this paper are not intrinsically different from the general population of GRBs.
If they are part of the same population, this triggers the question of whether our results (very low $\epsilon_B$, X-ray flux either below $\nu_c$ or suppressed by SSC, and prompt efficiencies around $\lesssim 20\%$) can be extended to the whole population or not.
Of course, since the requirement of LAT detection selects only the brightest bursts, the kinetic energies involved in their afterglow blast waves  are not representative of the whole population. 

Concerning the tendency to have very low $\epsilon_B$ values, we note that there is no reason to imagine a bias towards small values introduced by the study of a sample of bright bursts. The selection criterion is based on the detection in an energy range ($\sim$GeV) that lies above $\nu_c$, where the flux is very weakly dependent on $\epsilon_B$: $F_\nu\propto \epsilon_B^{(p-2)/2}$. This implies that if a selection bias exists, it goes in the opposite direction, since the GeV afterglow flux would be larger for higher values of $\epsilon_B$. Moreover, two GRBs with the same kinetic energy but very different $\epsilon_B$ would have very similar GeV fluxes, and then the same probability to be detected by LAT. 
We also note that small values of $\epsilon_B$ have been derived also in recent studies of larger samples of bursts without GeV radiation (\citealt{RBD(2014),Santana(2014),Zhang(2015),Wang(2015)}; see also the exercise of not including GeV data in the afterglow modelling of bursts detected in the GeV in \citealt{KBD(2009),KBD(2010),RBD(2011),Lemoine(2013)}).
It is not easy to understand why recent works (with and without LAT data) are highlighting the existence of $\epsilon_B$ values much smaller that those derived in previous analysis.
We can speculate that, at least in cases where data are not sufficient to break the degeneracy between the several free parameters (especially $\epsilon_B$ and $n$) solutions corresponding to very small $\epsilon_B$ values are usually discarded, in favour of values considered more standard.
Moreover, also in those cases where the free parameters are all constrained by the data, addition of GeV data obviously brings some new information and opens a region of the parameter space which is usually not investigated when high-energy observation are missing.
In fact, as we have proved in this work, GeV data are inconsistent with being a simple extrapolation of the X-ray data, and this rules out large $\epsilon_B$ values, in favour of smaller values. Small values of $\epsilon_B$ require a proper treatment of the IC cooling and KN effects. These effects modify the shape of the synchrotron spectrum and allow us to recover a satisfactorily modelling of the broad band data and to explain the apparent inconsistency between X-ray and GeV fluxes.
Our results, together with results from samples without LAT detection, suggest that the distribution of $\epsilon_B$ in the forward external shocks of GRBs is wider than what is usually believed, possibly rangingall the way down to $10^{-7}$.

The availability of LAT data affects not only the estimate of $\epsilon_B$, but also the estimates of the blast wave kinetic energy (and then, the estimates of the prompt efficiency). Also in this case, it is interesting to compare estimates derived by including and excluding LAT data. In our sample, the efficiencies derived from X-ray alone agree with results found for other bursts, i.e. large values close to unity (\S\ref{efficiency}). Again, only the inclusion of LAT data forces the modelling to prefer higher values of the kinetic energy, lowering the efficiency requirement on the prompt mechanism. The fact that the GeV detected bursts, which are most energetic indicate low efficiency during the prompt phase, suggests that the results are generic. Otherwise why would the brightest burst be also the least efficient?  

Regardless of whether or not these results can be extended to a large sample of bursts, our study outlines the need to perform broad band afterglow modeling in order to properly infer important properties like the energetics and the efficiency, and put constraints on the nature of the emission mechanisms in GRBs.

\section*{Acknowledgments}
We thank Pawan Kumar and Rodolfo Santana for helpful comments.
This work made use of data supplied by the UK Swift Science Data Centre at the University of Leicester.
The work was supported  by the ERC grant GRBs, by a grant from the Israel ISF - China NSF collaboration, by a grant from the Israel Space Agency, and by the I-Core Center of Excellence in Astrophysics. 

LN was supported by a Marie Curie Intra-European Fellowship of the European Community’s 7th Framework Programme (PIEF-GA-2013- 627715).

\appendix
\numberwithin{equation}{section}
\section[A]{SSC Compton parameter in the Thomson regime}
\label{ComptonY}
We give a short derivation of Eq. \ref{Yx} used for estimating the SSC Compton parameter in the Thomson regime.
We assume $2<p<3$ and $\nu_m<\nu_c$ which is the most relevant situation for GRB afterglows at times between $10^3-10^5$sec,
when late GeV and late X-ray / optical observations are taken.
Assuming an initial power law spectrum of electrons with a slope $p$, and taking into account evolution of the particle
spectrum due to cooling we write the density per unit energy as:
\begin{equation}
 \frac{dn_e'}{d\gamma_e}=
\left\{
  \begin{array}{l l}
    C (\gamma_e/\gamma_m)^{-p}, & \quad \gamma_m<\gamma_e<\gamma_c\\
    C (\gamma_c/\gamma_m)^{-p} (\gamma_e/\gamma_c)^{-p-1}, & \quad \gamma_c<\gamma_e\\
  \end{array} \right.
\end{equation}
where here and throughout this section, primes denote quantities in the comoving frame.
$C$ is given by the total density of electrons, $n_0'$:
\begin{equation}
\label{denscom}
n_0'=\int_{\gamma_m}^\infty \frac{dn_e'}{d\gamma_e}d\gamma_e =\frac{C}{p-1} (\gamma_m-\gamma_c^{-p+1}\gamma_m^p)+\frac{C\gamma_c}{p}\bigg(\frac{\gamma_c}{\gamma_m}\bigg)^{-p}.
\end{equation}
 Next, we relate $\gamma_m, \gamma_c$ to $\epsilon_e, \epsilon_B$:
\begin{equation}
\label{gammam}
\gamma_m=\frac{u\epsilon_e(p-2)}{m_e c^2 n_0(p-1)}, 
\end{equation}
\begin{equation}
\label{gammac}
\gamma_c=\frac{\gamma_{c,syn}}{1+Y}=\frac{6 \pi m_e c}{t_{dyn}' \sigma_T B^2(1+Y)}=\frac{3 m_e c^2}{4u(1+Y) \epsilon_B \Delta R' \sigma_T}, 
\end{equation}
where $u=\Gamma n_0' m_p c^2$ is the energy density of the shocked fluid, $t_{dyn}'$ is the time since the shock began
and $\Delta R'$ is the shell's width. The Compton $Y$ parameter in Eq. \ref{gammac} is actually $Y_c$,
which corresponds to electrons radiating at $\nu_c$; however, for all electrons in the Thomson regime, $Y$ is the same.

For $\gamma_c\gg\gamma_m$, Eq.\ref{denscom} yields $n_0'\approx C\gamma_m/(p-1)$ and we can write the average Lorentz factor squared as:
\begin{equation}
 \langle \gamma_e^2\rangle=\frac{\int_{\gamma_m}^\infty \frac{dn_e'}{d\gamma_e}\gamma_e^2d\gamma_e}{\int_{\gamma_m}^\infty \frac{dn_e'}{d\gamma_e}d\gamma_e}=\frac{1}{n_0'}\int_{\gamma_m}^\infty \frac{dn_e'}{d\gamma_e}\gamma_e^2d\gamma_e=\frac{(p-1) \gamma_m^{p-1} \gamma_c^{3-p}}{(3-p)(p-2)}.
\end{equation}
The Compton parameter is given by:
\begin{equation}
 Y=\frac{4}{3}\tau_e \langle \gamma_e^2\rangle=\frac{4(p-1)\tau_e \gamma_m^{p-1} \gamma_c^{3-p}}{3(3-p)(p-2)}
\end{equation}
Plugging in $\tau_e=n_0'\sigma_T \Delta R'$, and using Eqs. \ref{gammam}, \ref{gammac} we obtain:
\begin{equation}
\label{eq:Y}
 Y=\frac{\epsilon_e}{\epsilon_B (3-p) (1+Y)}\bigg(\frac{\gamma_m}{\gamma_c}\bigg)^{p-2}=\frac{\epsilon_e}{\epsilon_B (3-p) (1+Y)}\bigg(\frac{\nu_m}{\nu_c}\bigg)^{p-2\over 2}.
\end{equation}
Notice that in this equation, $Y$ appears also indirectly through $\nu_c$ (see Eq. \ref{eq:nuc}).
\end{document}